\documentclass[a4paper]{ustthesis}
\usepackage[square,numbers]{natbib}
\usepackage[T1]{fontenc}
\usepackage{longtable}
\usepackage{url}
\usepackage{multirow}
\usepackage{graphics}
\usepackage{graphicx}
\usepackage{colortbl}
\usepackage{xcolor}
\usepackage{times}
\usepackage{array}
\usepackage{listings}
\usepackage{amssymb}
\usepackage{subcaption}
\usepackage{listings}
\usepackage{adjustbox}
\usepackage{makecell}

\usepackage{tikz}
\usetikzlibrary{arrows}
\usetikzlibrary{positioning}
\newcommand{\oseries}[2]{#1 \otimes #2}
\newcommand{\oparal}[2]{#1 \oplus  #2}
\newcommand{\oloop}[1]{#1^\circledast}
\newcommand{\cfg}[1]{\textsf{cfg}(#1)}
\newcommand{\lt}[1]{\textsf{lt}(#1)}
\newcommand{\opt}[1]{\textsc{Opt}[#1]}
\newcommand{\vars}{\mathbb{V}}
\newcommand{\cost}{\textsc{Cost}}
\newcommand{\DP}{\texttt{dp}}
\newcommand{\compatible}{\leftrightharpoons}
\usepackage{amsmath}
\usepackage{cases}

\DeclareMathOperator*{\argmin}{arg\,min}

\tikzset{
    arrow/.style={thick,->,>=stealth, draw=blue!20!black},
    vertex/.style={rectangle, minimum width=10mm, minimum height=10mm, fill=green!10, draw=black!50},
    whitevertex/.style={rectangle, minimum width=10mm, minimum height=10mm, fill=white, draw=black!50},
    grayvertex/.style={rectangle, minimum width=10mm, minimum height=10mm, fill=gray!20, draw=gray!80},
    redvertex/.style={rectangle, minimum width=10mm, minimum height=10mm, fill=orange!20, draw=orange!80},
    bluevertex/.style={rectangle, minimum width=10mm, minimum height=10mm, fill=green!30, draw=green!80},
	vertexs/.style={thick, rectangle, minimum size=10mm, fill=blue!20, draw=black},
    vertext/.style={thick, rectangle, minimum size=10mm, fill=red!20, draw=black},
    vertexc/.style={thick, rectangle, minimum size=10mm, fill=green!10, draw=black},
    vertexb/.style={thick, rectangle, minimum size=10mm, fill=black, draw=black, text=white},
    vertexla/.style={minimum size=5mm, draw=none, text=red},
    emptynode/.style={thick, rectangle, minimum size=10mm, fill=white, draw=white, text=white},
    rect/.style={thick, rectangle, minimum size=10mm, fill=black!5, text=black}
}

\usepackage[colorlinks=true, allcolors=blue]{hyperref}

\usepackage{pdflscape}
\title{Enhancing Compiler Optimization Efficiency through Grammatical Decompositions of Control-Flow Graphs}  % Title of the thesis.
\author{CAI, Xuran}     % Author of the thesis.
\degree{\MPhil}             % Degree for which the thesis is. Options: \AM \MSc \MPhil \PhD
\stage{\Thesis}              % Stage of PhD document; use \Thesis for all other degree. Options: \PQE \Proposal \Thesis
\subject{Computer Science and Engineering} % Subject of the Degree.
\department{Department of Computer Science and Engineering}       % Department to which the thesis is submitted.
\advisor{Prof. Amir Kafshdar Goharshady, Department of Computer Science, University of Oxford
\newline
}     % Supervisor. Additional co-supervisor can be added using \member
\coadvisor{Prof. Jiasi Shen, Department of Computer Science and Engineering, HKUST}
\depthead{Prof. Xiaofang Zhou, Head of the Department of Computer Science and Engineering, HKUST}     % department head.
\defencedate{2025}{07}{16}     % \defencedate{year}{month}{day}.

\begin{document}
%\begin{CJK}{UTF8}{song}  % Bitstream Cyber Bit song ti

%\begin{CJK*}{UTF8}{gbsn} % Arphic song ti

%%%%%%%%%%%%%%%%%%%%%%%%%%%%%%%%%%%%%%%%%%%%%%%%%%%%%%%%%%%%%%%%%%%%%%%%%
%                                                                       %
% Now the actual Thesis. The order of output MUST be followed:          %
%                                                                       %
%    1) TITLEPAGE                                                       %
%                                                                       %
% The \maketitle command generates the Title page as well as the        %
% Signature page.                                                       %
%                                                                       %
%%%%%%%%%%%%%%%%%%%%%%%%%%%%%%%%%%%%%%%%%%%%%%%%%%%%%%%%%%%%%%%%%%%%%%%%%

\maketitle

%%%%%%%%%%%%%%%%%%%%%%%%%%%%%%%%%%%%%%%%%%%%%%%%%%%%%%%%%%%%%%%%%%%%%%%%%
%                                                                       %
%     3) ACKNOWLEDGMENTS                                                %
%                                                                       %
% \acknowledgments and \endacknowledgments defines the                  %
% Acknowledgments of the author of the Thesis.                          %
%                                                                       %
%%%%%%%%%%%%%%%%%%%%%%%%%%%%%%%%%%%%%%%%%%%%%%%%%%%%%%%%%%%%%%%%%%%%%%%%%

\acknowledgments
I would like to express my heartfelt gratitude to those who have supported me throughout my journey to complete this thesis.

First and foremost, I would like to thank Professor Amir Goharshady for guiding me into the research area and providing invaluable support and suggestions. Your mentorship has been instrumental in shaping my research and academic growth. I am also deeply grateful to Professor Jiasi Shen for her kind agreement to be my HKUST co-supervisor.

I am also deeply grateful to Professor Sunil Arya,  Professor Yi Ke, and Professor Lionel Parreaux for their kind support during my MPhil experience. Your encouragement and insights have greatly enriched my understanding and have been a source of motivation.

A special thanks to all the members of the ALPACAS research group. I feel fortunate to have joined such a creative and kind group of individuals. Your collaboration and camaraderie have made this journey enjoyable and fulfilling.

I would like to extend my heartfelt appreciation to my family for their unwavering support during challenging times. Additionally, I am grateful to my girlfriend, Trinity, for her accompaniment and encouragement, which have helped me through the downs of this journey.

Lastly, I would like to acknowledge all my old and new friends. Your companionship and encouragement have made this experience memorable.

Thank you all for your contributions to my academic and personal journey.
\endacknowledgments
\newpage

%%%%%%%%%%%%%%%%%%%%%%%%%%%%%%%%%%%%%%%%%%%%%%%%%%%%%%%%%%%%%%%%%%%%%%%%%
%                                                                       %
%     4) TABLE OF CONTENTS                                              %
%                                                                       %
%%%%%%%%%%%%%%%%%%%%%%%%%%%%%%%%%%%%%%%%%%%%%%%%%%%%%%%%%%%%%%%%%%%%%%%%%

\tableofcontents

%%%%%%%%%%%%%%%%%%%%%%%%%%%%%%%%%%%%%%%%%%%%%%%%%%%%%%%%%%%%%%%%%%%%%%%%%
%                                                                       %
%     5) LIST OF FIGURES (If Any)                                       %
%                                                                       %
%%%%%%%%%%%%%%%%%%%%%%%%%%%%%%%%%%%%%%%%%%%%%%%%%%%%%%%%%%%%%%%%%%%%%%%%%

\listoffigures

%%%%%%%%%%%%%%%%%%%%%%%%%%%%%%%%%%%%%%%%%%%%%%%%%%%%%%%%%%%%%%%%%%%%%%%%%
%                                                                       %
%     6) LIST OF TABLES (If Any)
%                                                                       %
%%%%%%%%%%%%%%%%%%%%%%%%%%%%%%%%%%%%%%%%%%%%%%%%%%%%%%%%%%%%%%%%%%%%%%%%%

%\listoftables
\newpage
  \pchapter{Publications}

\begin{itemize}
    \item X. Cai, A.K. Goharshady, S. Hitarth, C.K. Lam \\
\textbf{Faster Chaitin-like Register Allocation via Grammatical Decompositions of Control-Flow Graphs} \cite{RegisterAllocation}\\
International Conference on Architectural Support for Programming Languages and Operating Systems, ASPLOS 2025

\item X. Cai, A.K. Goharshady \\
\textbf{Faster lifetime-optimal speculative partial redundancy elimination for goto-free programs} \cite{cai2024faster}\\
International Symposium on Software Engineering: Theories, Tools, and Applications, SETTA 2024
\end{itemize}

%%%%%%%%%%%%%%%%%%%%%%%%%%%%%%%%%%%%%%%%%%%%%%%%%%%%%%%%%%%%%%%%%%%%%%%%%
%                                                                       %
%     8) The Actual Contents                                            %
%                                                                       %
% The command \chapters MUST BE USED to ensure that the entire content  %
% of the Thesis is double-spaced (in version 1.0).                      %
%                                                                       %
% However, in version 2.0, \chapters will be automatically added in     %
% the beginning of the first chapter.                                   %
%                                                                       %
%%%%%%%%%%%%%%%%%%%%%%%%%%%%%%%%%%%%%%%%%%%%%%%%%%%%%%%%%%%%%%%%%%%%%%%%%

%%\chapters         % Not necessary with ustthesis.cls (v2.0).

%%%%%%%%%%%%%%%%%%%%%%%%%%%%%%%%%%%%%%%%%%%%%%%%%%%%%%%%%%%%%%%%%%%%%%%%%
%                                                                       %
% Each chapter is defined via the \chapter command. The usual sectional %
% commands of LaTeX are also available.                                 %
%                                                                       %
%%%%%%%%%%%%%%%%%%%%%%%%%%%%%%%%%%%%%%%%%%%%%%%%%%%%%%%%%%%%%%%%%%%%%%%%%

\chapter{Introduction}
\label{chp_intro}
\paragraph{Compiler Optimization.} Compiler optimization \cite{AhoCompilers2006} is the process of enhancing the efficiency of software applications by transforming code during the compilation phase. This involves altering a program's source code or its intermediate representation to improve performance metrics such as execution speed, memory usage, and energy consumption. Various compiler optimization tasks must be addressed to meet different objectives and accommodate diverse architectures.

The significance of compiler optimization is immense, as it directly influences the runtime performance of applications. As software complexity increases and the demand for resource-efficient solutions grows, effective optimization techniques can result in faster execution, lower resource consumption, and an enhanced overall user experience. Furthermore, optimizing code enables developers to create more efficient programs, which ultimately leads to improved software quality and system performance. As technology evolves, compiler optimization remains a vital area of research and development within computer science.

\paragraph{Structured Program.} In this thesis, when referring to a structured program or a goto-free structured program, we mean the program generated from the following grammar, which is similar to that of \cite{DBLP:journals/iandc/Thorup98}.

\begin{equation} \label{gram:prog}
    \begin{array}{rl}
    P := & \epsilon \mid \texttt{break} \mid \texttt{continue} \mid P ; P\\ 
         & \mid \texttt{if} ~\varphi~ \texttt{then} ~P~ \texttt{else} ~P~ \texttt{fi} \mid \texttt{while} ~\varphi~ \texttt{do} ~P~ \texttt{od}.
    \end{array}
\end{equation}

Here, \( \epsilon \) represents a neutral statement that does not affect control flow, such as a variable assignment, and \( \varphi \) is a boolean expression. We say a program \( P \) is \emph{closed} if every \texttt{break} and \texttt{continue} statement appears within the body of a \texttt{while} loop. The semantics of a program \( P \) will be defined in the usual manner. In this section, we focus solely on the control flow graph of a program \( P \). It is also worth noting that other common constructs, such as \texttt{for} loops and \texttt{switch} statements, can be defined as syntactic sugar \cite{}. Specifically, a \texttt{switch} statement with \( k \) jumps can be modeled as \( k \) \texttt{if} statements.

\paragraph{Graph Theory and Control Flow Graph.} Graph theory \cite{diestel2024graph}  is a branch of mathematics that studies the properties and relationships of graph structures made up of vertices (or nodes) connected by edges. Graphs are typically represented as \( G = (V, E) \), where \( G \) denotes the graph, \( V \) is the set of all vertices, and \( E \) is the set of all edges. This mathematical framework is particularly valuable in computer science for modeling and solving problems that can be represented as graphs.

In the realm of programming and compilers, control flow graphs (CFGs) \cite{allen1970control} are a specific type of directed graph that illustrates the flow of a program's execution. In a CFG, nodes represent basic blocks of code—sequences of instructions with a single entry and exit point—while directed edges indicate the possible control flow paths between these blocks. This structure provides a clear visualization of program execution, facilitating various optimization techniques.

Modern compilers, such as GCC \cite{gough2004introduction}, SDCC \cite{sdcc}, and LLVM \cite{sarda2015llvm}, leverage control flow graphs (CFGs) to analyze programs and generate efficient code. While certain properties of CFG families have been identified, there remains considerable scope for discovering more powerful algorithms for the compiler optimization tasks. 

\paragraph{PCSP.} 
Constraint Satisfaction Problems (CSPs) and constraint-based reasoning have emerged as effective methodologies for problem-solving. These approaches involve determining values for variables while adhering to constraints that define permissible combinations of these values\cite{PCS}. Many common graph-related problems, such as the graph coloring problem, can be formulated as CSPs.

However, there are instances where it may be infeasible or impractical to find complete solutions to these problems. In such cases, we may aim for partial solutions, specifically by satisfying the maximum number of constraints while minimizing costs. This thesis will concentrate on the binary relationships within Partial Constraint Satisfaction Problems (PCSPs), where all constraints involve only two variables\cite{Koster2002SolvingCS}.

PCSPs have a wide array of applications. A notable example that translates elegantly into the PCSP framework is the MAX-SAT problem\cite{KOSTER199889}. Additionally, various compiler optimization tasks, particularly those related to graph theory, can be represented as PCSPs. Examples include register allocation\cite{RegisterAllocation}, the LOSPRE\cite{cai2024faster}, and placement of bank selection instructions\cite{bank_selection}.

The NP-hardness of PCSPs with domain sizes of at least three is established through a reduction from the 3-coloring problem in graphs. Koster et al. \cite{KOSTER199889}demonstrated, via a reduction from MAX-SAT, that PCSPs become NP-hard even when all domains are restricted to size two. Computational experiments support these findings in practical scenarios. In Koster et al. \cite{KOSTER199889}, a polynomial approach was employed with limited success.

\paragraph{Contributions.} In this work, we introduce a novel decomposition for control flow graphs (CFGs) along with a general solution for the Partial Constraint Satisfaction Problem (PCSP) over CFGs. My contributions include:

\begin{itemize}
    \item We designed a new decomposition called SPL-decomposition, which precisely covers CFGs of goto-free structured programs.
    \item We developed a general parameterized algorithm for PCSP using SPL-decomposition, achieving a time complexity of \(O(|G|\cdots|D|^5)\), where $G$ is the CFG and $D$ is the domain size of the PCSP.
    \item We identified three specific compiler optimization tasks that can be formulated as PCSPs, enabling the application of my algorithm. These tasks include register allocation, Lifetime-optimal Speculative Partial Redundancy Elimination (LOSPRE), and the placement of bank selection instructions.
    \item We implemented these three compiler optimization tasks in SDCC and conducted regression tests, comparing the results with state-of-the-art algorithms. The results demonstrate that my algorithm significantly improves performance across all three tasks.
\end{itemize}

\paragraph{Limitations.} The main limitation of my decomposition and algorithm is that they are applicable only to goto-free structured programs. They do not extend to programs that utilize GOTO statements or non-structured constructs. Furthermore, the algorithm requires the control flow graph (CFG) to strictly follow the program's execution flow. If any optimizations modify the CFG structure, although such instances are uncommon, my algorithm would become ineffective.

\paragraph{Thesis Organization.} In Chapter 2, We provide background information on parameterized algorithms and tree decomposition. Chapter 3 introduces a novel decomposition method called SPL-decomposition, which is primarily based on our own paper \cite{RegisterAllocation}. Chapters 4 and 5 focus on two compiler optimization tasks: Register Allocation\cite{RegisterAllocation} and LOSPRE\cite{cai2024faster}, demonstrating how they can be addressed using SPL-decomposition. In Chapter 6, we present the PCSP problem along with a general solution. Chapter 7 introduces another compiler optimization task, the Placement of Bank Selection, and explains how to encode it as a PCSP. Chapter 8 presents experimental results comparing our approach to previous state-of-the-art solutions, and finally, Chapter 9 concludes the discussion.

\newpage

\chapter{Background}
\label{chp_background}
\paragraph{Parameterized Algorithm.}
Parameterized algorithms \cite{cygan2015parameterized}  represent a class of algorithms designed to address computational problems by concentrating on specific parameters that can simplify the problem's complexity. Rather than analyzing the problem solely based on its input size, parameterized algorithms leverage additional parameters, allowing for more efficient solutions in certain cases. Many NP-hard problem in Computer Science area is solved by parameterized algorithm like \cite{gpar1, gpar2}.

A key concept in this domain is Fixed-Parameter Tractability (FPT) \cite{downey2013fundamentals}. An algorithm is deemed FPT if it can solve a problem within a time complexity of the form \(O(f(k) \cdot n^c)\), where \(f(k)\) is a function of a parameter \(k\) (typically much smaller than \(n\)), and \(c\) is a constant. This implies that the algorithm's running time is primarily influenced by the parameter \(k\), making it feasible to solve problems that would otherwise be intractable for large inputs. 

Considering the minimum vertex cover problem\cite{khuller2002algorithms}, a classic problem in graph theory and computer science. It involves finding the smallest subset of vertices in a given graph such that every edge in the graph is incident to at least one vertex in this subset. In other words, a vertex cover is a set of vertices that "covers" all the edges of the graph.

This is a typical NP-hard problem; however, if we reframe the question to ask whether it is possible to find a subset of the vertex cover with a size of at most \( k \), we can use \( k \) as our parameter. This allows us to develop a fixed-parameter tractable (FPT) algorithm to solve the problem efficiently using the following approach:

\begin{lstlisting}
# FPT algorithm for Vertex Cover
def vertex_cover(graph, k):
    if k < 0:
        return None  # Not possible
    if graph.isEmpty():
        return set()  # No edges left

    # Choose an arbitrary edge (u, v)
    u, v = graph.edges[0]
    # Create two branches: include u or include v in the cover
    cover_with_u = vertex_cover(graph.remove_vertex(u), k - 1)
    cover_with_v = vertex_cover(graph.remove_vertex(v), k - 1)

    if cover_with_u is not None:
        return cover_with_u.union({u})
    if cover_with_v is not None:
        return cover_with_v.union({v})

    return None  # No valid cover found
\end{lstlisting}

As each recursive call takes constant time and the recursion has a maximum depth of \( k \), with each node having two branches, the time complexity of this algorithm is \( O(2^k \cdot n) \). In this scenario, \( f(k) = 2^k \), which confirms that this is a fixed-parameter tractable (FPT) algorithm.

Another important class of parameterized algorithms is the XP (Exponential Time Parameterized) algorithm \cite{downey2013fundamentals}. An XP algorithm exhibits a running time of \(O(n^{f(k)})\), where \(k\) is a parameter. While XP algorithms may not be as efficient as FPT algorithms, they can still provide practical solutions for problems where the parameter \(k\) is small relative to the input size \(n\).

There is also an Exponential Time Parameterized (XP) algorithm for the edited vertex cover problem. Since we are interested in finding a \( k \)-subset vertex cover, we can simply enumerate all possible \( k \)-subsets and check if any of them is valid. In this approach, there are $ \binom{n}{k} $ possible \( k \)-subsets, and verifying the validity of each vertex cover requires at most \( O(n) \) time. Consequently, the overall time complexity of this algorithm is \( O(n^{k+1}) \), where \( f(k) = k + 1 \) in this case.

By harnessing the power of parameterized algorithms, We can develop more effective solutions for complex problems in compiler optimization and model checking, potentially resolving some NP-hard problems in linear time and ultimately enhancing the performance of modern programming languages.

\paragraph{Tree-decomposition and Treewidth.}

    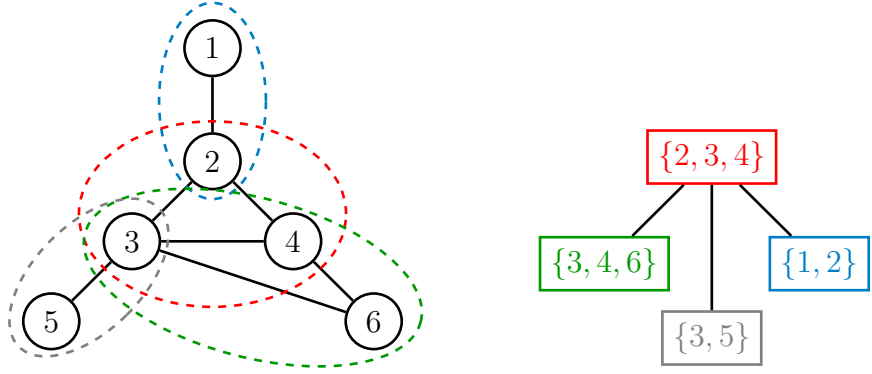
\begin{figure*}
\begin{center}
	$\begin{matrix}
		\begin{tikzpicture}[node distance={15mm}, line width=1pt, main/.style = {draw, circle}] 
			\node[main] (1) at (0,0){$1$}; 
			\node[main] (2) [below of=1]{$2$}; 
			\node[main] (3) [below left of=2]{$3$}; 
			\node[main] (4) [below right of=2]{$4$}; 
			\node[main] (5) [below left of=3]{$5$}; 
			\node[main] (6) [below right of=4]{$6$}; 
			\draw [] (1) -- (2); 
			\draw [] (2) -- (3); 
			\draw [] (2) -- (4); 
			\draw [] (3) -- (4); 
			\draw [] (3) -- (5); 
			\draw [] (3) -- (6); 
			\draw [] (4) -- (6); 
			
			\node[text width=0cm, red] at (2,0){$ $};
			\draw[rotate=0,line width=1pt,dashed,cyan!60!blue] (0,-0.7) ellipse (20pt and 37pt);
			
			\node[text width=0cm, red] at (4.8,-1.8){$ $};
			\draw[rotate=0,line width=1pt,dashed,red] (0,-2.2) ellipse (50pt and 35pt);
			
			\node[text width=0cm, red] at (5.8,-3.6){$ $};
			\draw[rotate=-15,line width=1pt,dashed,green!60!black] (1.3,-2.8) ellipse (65pt and 30pt);
			
			\node[text width=0cm, red] at (-0.1,-3.6){$ $};
			\draw[rotate=-45,line width=1pt,dashed,gray] (1,-3.3) ellipse (20pt and 37pt);
		\end{tikzpicture}
		&   \hspace{-2cm}
		\begin{tikzpicture}[node distance={20mm}, line width=1pt, main/.style = {draw, rectangle}] 
			\node[main] (1) [red]{$\{2, 3, 4\}$}; 
			\node[main] (2) [cyan!60!blue, below right of=1]{$\{1, 2\}$}; 
			\node[main] (3) [below of=1,node distance=24mm, gray]{$\{3, 5\}$}; 
			\node[main] (4) [green!60!black, below left of=1]{$\{3, 4, 6\}$}; 
			\draw [] (1) -- (2); 
			\draw [] (1) -- (3);
			\draw [] (1) -- (4);
		\end{tikzpicture}
	\end{matrix}$
 \end{center}
	\caption{A graph $G$ (left) and a tree decomposition of $G$ (right). }
	\label{fig:tw}
\end{figure*}

A tree decomposition \cite{robertson1986graph}  of a graph \( G = (V, E) \) is a pair \( (T, \{B_t\}_{t \in T}) \), where \( T \) is a tree and each node \( t \in T \) is associated with a bag \( B_t \subseteq V \). Figure~\ref{fig:tw} is a tree decomposition example from \cite{conradof2023bounded}. A tree decomposition satisfies three conditions: 

\begin{itemize}
    \item \textbf{Covering Condition}: For every vertex \( v \in V \), there exists at least one node \( t \in T \) such that \( v \in B_t \). 
    \item \textbf{Edge Condition}: For every edge \( (u, v) \in E \), there exists a node \( t \in T \) such that both \( u \) and \( v \) are contained in the bag \( B_t \). 
    \item \textbf{Connectivity Condition}: For every vertex \( v \in V \), the nodes \( t \in T \) that contain \( v \) form a connected subtree of \( T \). 
\end{itemize}
The treewidth of a graph \( G \) is defined as the minimum width of all possible tree decompositions of \( G \), where the width is the size of the largest bag minus one.

Tree decomposition is particularly valuable for solving NP-hard problems, as it enables the effective application of dynamic programming techniques. By capitalizing on the tree's structure, algorithms can operate on smaller, more manageable components of the graph, leading to more efficient solutions. This approach is widely utilized in areas such as graph algorithms, network design, and, importantly, in compiler optimization, where it can aid in analyzing and optimizing program structures.

Recent research has demonstrated that all goto-free structured control flow graphs (CFGs) have a treewidth of at most 7 \cite{DBLP:journals/iandc/Thorup98}. This property enables the application of tree decomposition for analyzing CFGs, treating the treewidth as a constant parameter, which can be leveraged to generate parameterized algorithms. By taking advantage of this characteristic, we can develop efficient algorithms for various optimization tasks in compilers like \cite{gtw1, gtw2}.

One notable application of tree decomposition is in dynamic programming, particularly for solving problems that may be computationally intensive on general graphs \cite{de1997algorithms}. Consider a graph problem \( P \) on a graph \( G = (V, E) \), and let \( (T, \{B_t\}_{t \in T}) \) be a tree decomposition of \( G \). For each node \( t \) in the tree, let \( G_t \) be the induced subgraph with vertices in \( B_t \). Suppose \( S_t \) is a table that contains the information necessary to solve problem \( P \). If \( S_t \) satisfies the following properties:
\begin{enumerate}
    \item For each graph \( G_t \), problem \( P \) can be solved solely using table \( S_t \).
    \item For each \( t \) that is a leaf node in \( T \), \( S_t \) can be computed exclusively based on \( G_t \).
    \item \( S_t \) can be calculated using \( G_t \) and the tables of \( t \)'s children in the tree.
\end{enumerate}

With these three properties, we can efficiently execute dynamic programming from the bottom of the tree to the top. Due to the small treewidth, the size of \( |G_t| \) is constrained, allowing the processing at each \( G_t \) to be completed in constant time. This results in a linear parameterized algorithm. In Chapter 4, I will demonstrate that the PCSP adheres to all three properties, thus enabling an efficient linear time complexity algorithm for PCSP over control flow graphs (CFGs).

\newpage

\chapter{SPL-decomposition}
\label{chp_background}
In this chapter, we will provide a detailed introduction to SPL-decompositions, following the insights presented in my own papers \cite{RegisterAllocation,cai2024faster}.

An SPL graph $G = (V, E, S, T, B, C)$ is a directed graph $(V, E)$ with four distinct special nodes $S, T, B, C \in V,$ which are respectively called the \emph{start}, \emph{terminate}, \emph{break} and \emph{continue} nodes, generated by the grammar below:
\begin{equation} \label{gram:spl}
	\begin{array}{rl}
		G := & A_\epsilon ~|~ A_{\texttt{break}} ~|~ A_{\texttt{continue}} ~|~ \oseries{G}{G} ~|~ \oparal{G}{G} ~|~ \oloop{G}
	\end{array}
\end{equation}
We now explain the operations in this grammar.

\begin{figure}
	\begin{center}
	\begin{tabular}{cc|ccc|cc}
		\begin{tikzpicture}[scale=0.6]
			\node[vertex] (S) {$S$};
			\node[vertex,fill=blue!20] (T)  [below=of S] {$T$};
			\node[vertex,fill=red!20] (C)  [right=of S] {$C$};
			\node[vertex,fill=cyan!20] (B)  [below=of C] {$B$};
			\draw[arrow] (S) -- (T) ;
		\end{tikzpicture} & \quad & \quad &
		\begin{tikzpicture}[scale=0.6]
			\node[vertex] (S) {$S$};
			\node[vertex,fill=blue!20] (T)  [below=of S] {$T$};
			\node[vertex,fill=red!20] (C)  [right=of S] {$C$};
			\node[vertex,fill=cyan!20] (B)  [below=of C] {$B$};
			\draw[arrow] (S) -- (B) node [midway, above] {};
		\end{tikzpicture} & \quad & \quad &
		\begin{tikzpicture}[scale=0.6]
			\node[vertex] (S) {$S$};
			\node[vertex,fill=blue!20] (T)  [below=of S] {$T$};
			\node[vertex,fill=red!20] (C)  [right=of S] {$C$};
			\node[vertex,fill=cyan!20] (B)  [below=of C] {$B$};
			\draw[arrow] (S) -- (C) node [midway, above] {};
		\end{tikzpicture} 
	\end{tabular}
\end{center}
\caption{Atomic SPL graphs: $A_\epsilon$ (left), $A_{\texttt{break}}$ (middle), and $A_{\texttt{continue}}$ (right).}
\label{fig:atom}
\end{figure}
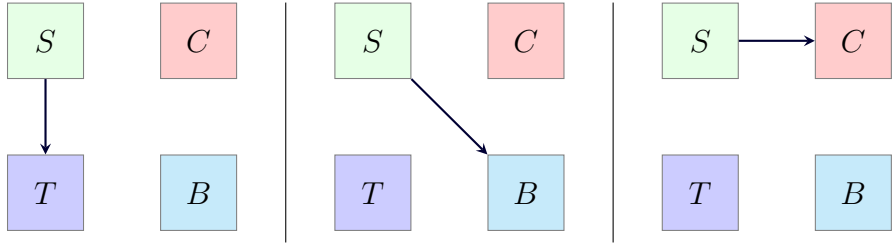

\paragraph{Atomic Node.}
There are three different atomic SPL graphs: $A_\epsilon$, $A_{\texttt{break}}$, and $A_{\texttt{continue}}$.
    All of them contain only the four special nodes and only one edge as shown in Figure~\ref{fig:atom}.

\begin{figure*}
	\begin{center}
		\begin{tabular}{cc|ccc|cc}
        \scalebox{0.7}{
			\begin{tikzpicture}[scale=0.6]
				\node[vertex] (S1) {$S_1$};
				\node[vertex,fill=blue!20] (T1)  [below=of S1] {$T_1$};
				\node[vertex,fill=red!20] (C1)  [right=of S1] {$C_1$};
				\node[vertex,fill=cyan!20] (B1)  [below=of C1] {$B_1$};
				\draw[arrow] (S1) -- (T1) node [midway, left] {};
				\draw[arrow] (S1) -- (B1) node [midway, above] {};
				\node[emptynode] (R1) [below=of B1] {};
				
				\node (z) [right=of B1] {\fontsize{52}{58}\sffamily\bfseries$\oseries{}{}$};
				
				\node[vertex] (S2) [right=of z] {$S_2$};
				\node[vertex,fill=blue!20] (T2)  [below=of S2] {$T_2$};
				\node[vertex,fill=red!20] (C2)  [right=of S2] {$C_2$};
				\node[vertex,fill=cyan!20] (B2)  [below=of C2] {$B_2$};
				\draw[arrow] (S2) -- (T2) ;
				
				\node (e) [right=of C2] {\fontsize{52}{58}\sffamily\bfseries$=$};
				
				\node[vertex,fill=yellow!20,dashed] (M) [right=of e]{$M$};
				\node[vertex] (S) [above=of M] {$S_1$};
				\node[vertex,fill=blue!20] (T)  [below=of M] {$T_2$};
				\node[vertex,fill=red!20,dashed] (C)  [right=of S] {$C$};
				\node[vertex,fill=cyan!20,dashed] (B)  [below=of C] {$B$};
				\draw[arrow] (S) -- (M) ;
				\draw[arrow] (S) -- (B) ;
				\draw[arrow] (M) -- (T) ;
			\end{tikzpicture}} \\ \\ \hline \hline \\
			\scalebox{0.7}{\begin{tikzpicture}[scale=0.6]
				\node[vertex] (S1) {$S_1$};
				\node[vertex,fill=blue!20] (T1)  [below=of S1] {$T_1$};
				\node[vertex,fill=red!20] (C1)  [right=of S1] {$C_1$};
				\node[vertex,fill=cyan!20] (B1)  [below=of C1] {$B_1$};
				\draw[arrow] (S1) -- (T1) ;
				\node[emptynode] (R1) [below=of B1] {};
				
				\node (z) [right=of B1] {\fontsize{52}{58}\sffamily\bfseries$\oseries{}{}$};
				
				\node[vertex] (S2) [right=of z] {$S_2$};
				\node[vertex,fill=blue!20] (T2)  [below=of S2] {$T_2$};
				\node[vertex,fill=red!20] (C2)  [right=of S2] {$C_2$};
				\node[vertex,fill=cyan!20] (B2)  [below=of C2] {$B_2$};
				\draw[arrow] (S2) -- (T2) node [midway, left] {};
				\draw[arrow] (S2) -- (C2) node [midway, above] {};
				
				\node (e) [right=of C2] {\fontsize{52}{58}\sffamily\bfseries$=$};
				
				\node[vertex,fill=yellow!20,dashed] (M)  [right=of e] {$M$};
				\node[vertex] (S) [above=of M]{$S_1$};
				\node[vertex,fill=blue!20] (T)  [below=of M] {$T_2$};
				\node[vertex,fill=red!20,dashed] (C)  [right=of M] {$C$};
				\node[vertex,fill=cyan!20,dashed] (B)  [below=of C] {$B$};
				\draw[arrow] (S) -- (M);
				\draw[arrow] (M) -- (T);
				\draw[arrow] (M) -- (C);
			\end{tikzpicture}
			}
		\end{tabular}
	\end{center}
	\caption{Two examples of the series operation $\oseries{}{}$. }
	\label{fig:series}
\end{figure*}
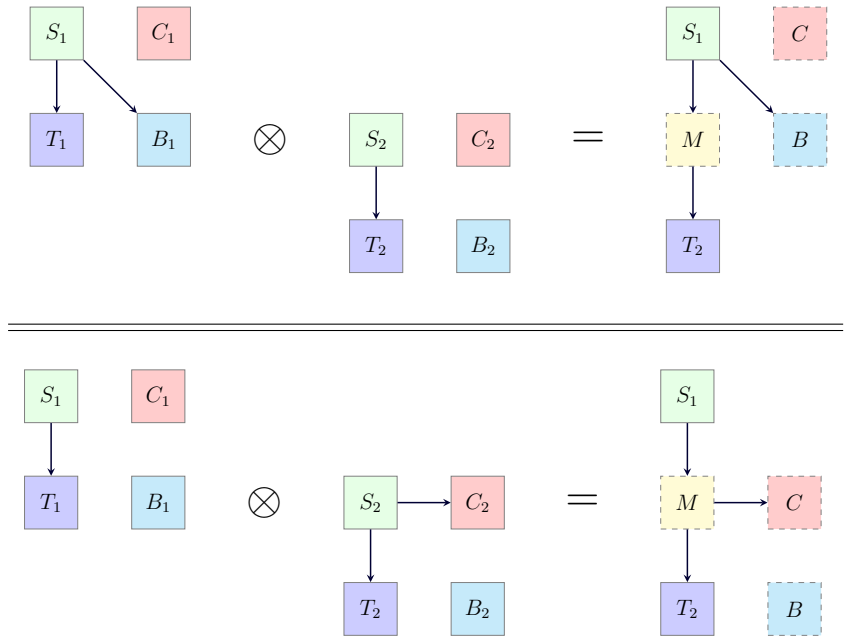

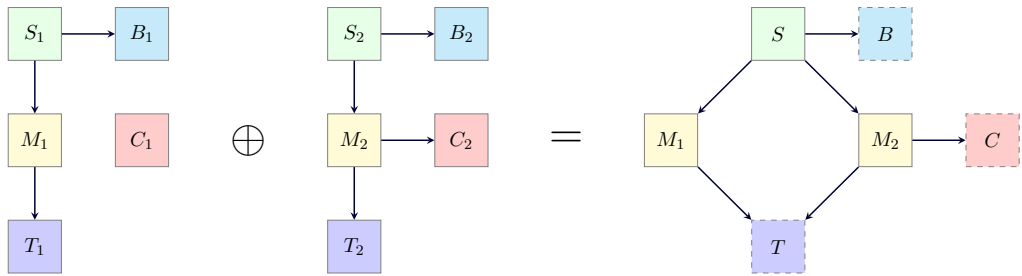
\begin{figure*}
\scalebox{0.7}{
    \begin{tikzpicture}[scale=0.6] % Adjusted scale to make the figure smaller
        \node[vertex,fill=yellow!20] (M1) {\small $M_1$}; % Reduced font size
        \node[vertex] (S1) [above=of M1] {\small $S_1$};
        \node[vertex,fill=blue!20] (T1)  [below=of M1] {\small $T_1$};
        \node[vertex,fill=cyan!20] (B1)  [right=of S1] {\small $B_1$};
        \node[vertex,fill=red!20] (C1)  [below=of B1] {\small $C_1$};
        \draw[arrow] (S1) -- (M1) ;
        \draw[arrow] (S1) -- (B1) ;
        \draw[arrow] (M1) -- (T1) ;

        \node (z) [right=of C1] {\fontsize{36}{40}\sffamily\bfseries$\oparal{}{}$}; % Adjusted font size

        \node[vertex,fill=yellow!20] (M2) [right=of z]  {\small $M_2$};
        \node[vertex=] (S2) [above=of M2]{\small $S_2$};
        \node[vertex,fill=blue!20] (T2)  [below=of M2] {\small $T_2$};
        \node[vertex,fill=red!20] (C2)  [right=of M2] {\small $C_2$};
        \node[vertex,fill=cyan!20] (B2)  [above=of C2] {\small $B_2$};
        \draw[arrow] (S2) -- (M2);
        \draw[arrow] (S2) -- (B2);
        \draw[arrow] (M2) -- (T2);
        \draw[arrow] (M2) -- (C2);

        \node (e) [right=of C2] {\fontsize{36}{40}\sffamily\bfseries$=$}; % Adjusted font size

        \node[vertex,fill=yellow!20] (aM1) [right=of e] {\small $M_1$};
        \node[vertex] (aS) [above right=of aM1] {\small $S$};
        \node[vertex,fill=yellow!20] (aM2) [below right=of aS] {\small $M_2$};
        \node[vertex,dashed,fill=blue!20] (aT)  [below left=of aM2] {\small $T$};
        \node[vertex,dashed,fill=red!20] (aC)  [right=of aM2] {\small $C$};
        \node[vertex,dashed,fill=cyan!20] (aB)  [right=of aS] {\small $B$};
        \draw[arrow] (aS) -- (aM1);
        \draw[arrow] (aS) -- (aM2);
        \draw[arrow] (aS) -- (aB);
        \draw[arrow] (aM1) -- (aT);
        \draw[arrow] (aM2) -- (aT);
        \draw[arrow] (aM2) -- (aC);
    \end{tikzpicture}}
    \caption{An example of the parallel operation $\oparal{}{}$. }
    \label{fig:parallel}
\end{figure*}

\paragraph{SPL Operations.}	SPL defines three operations. Let $G_1 = (V_1, E_1, S_1, T_1,$ $B_1, C_1)$ and 
	$G_2 = (V_2, E_2, S_2, T_2, B_2, C_2)$ be two disjoint SPL graphs. Then, the graphs obtained by the following operations are also SPL graphs.
\begin{itemize}
	\item \textbf{Series Operation.} $\oseries{G_1}{G_2}$ is generated by taking the union of $G_1$ and $G_2$ and merging the pairs of vertices $M = (T_1, S_2)$,  $B = (B_1, B_2)$, and $C = (C_1, C_2)$. The distinguished vertices of $\oseries{G_1}{G_2}$ are $(S_1, T_2, B, C)$. It is easy to verify that the series operation is associative. Figure~\ref{fig:series} shows two examples of the series operation.
	
	\item \textbf{Parallel Operation.} $\oparal{G_1}{G_2}$ is generated by taking union of $G_1$ and $G_2$ and 
	merging the pairs of vertices $S = (S_1, S_2)$, $T = (T_1, T_2)$, $B = (B_1, B_2)$, and $C = (C_1, C_2)$. The 
	special vertex tuple of $\oseries{G_1}{G_2}$ is $(S, T, B, C).$ Figure~\ref{fig:parallel} shows an example 
	of this operation.
		
	\item \textbf{Loop Operation.} $\oloop{G_1}$ is generated by adding four new vertices $S, T, B, C$ to $G_1$ and 
	then adding the following edges: $(S, S_1), (S, T), (T_1, S), (C_1, S),$ and $(B_1, T).$ The special vertex 
	tuple of $\oloop{G_1}$ is $(S, T, B, C).$ Figure~\ref{fig:loop} shows an example of the loop operation.

\end{itemize}

	\begin{figure*}
	\begin{center}
    \scalebox{0.7}{
		\begin{tikzpicture}[scale=0.6]
			\node[vertex,fill=yellow!20] (aM1)  {$M_1$};
			\node[vertex] (aS) [above right=of aM1] {$S_1$};
			\node[vertex,fill=yellow!20] (aM2) [below right=of aS] {$M_2$};
			\node[vertex,fill=blue!20] (aT)  [below left=of aM2] {$T_1$};
			\node[vertex,fill=red!20] (aC)  [above right=of aM2] {$C_1$};
			\node[vertex,fill=cyan!20] (aB)  [left=of aS] {$B_1$};
			\draw[arrow] (aS) -- (aM1);
			\draw[arrow] (aS) -- (aM2);
			\draw[arrow] (aS) -- (aB);
			\draw[arrow] (aM1) -- (aT);
			\draw[arrow] (aM2) -- (aT);
			\draw[arrow] (aM2) -- (aC);
			
			\node (star) [right=of aC] {\fontsize{52}{58}\sffamily\bfseries$\oloop{}$};
			
			\node (e) [below right=of star] {\fontsize{52}{58}\sffamily\bfseries$=$};
			
			\node[vertex,fill=yellow!20] (baM1) [right=of e] {$M_1$};
			\node[vertex,fill=yellow!20] (baS) [above right=of baM1] {$S_1$};
			\node[vertex,dashed] (S) [above right=of baS] {$S$};
			\node[vertex,fill=cyan!20,dashed] (B) [right=of S] {$B$};
			\node[vertex,fill=red!20,dashed] (C) [right=of B] {$C$};
			\node[vertex,fill=yellow!20] (baM2) [below right=of baS] {$M_2$};
			\node[vertex,fill=yellow!20] (baT)  [below left=of baM2] {$T_1$};
			\node[vertex,fill=yellow!20] (baC)  [above right=of baM2] {$C_1$};
			\node[vertex,fill=yellow!20] (baB)  [left=of baS] {$B_1$};
			\node[vertex,fill=blue!20,dashed] (T) [left=of S] {$T$};
			\draw[arrow] (baS) -- (baM1);
			\draw[arrow] (baS) -- (baM2);
			\draw[arrow] (baS) -- (baB);
			\draw[arrow] (baM1) -- (baT);
			\draw[arrow] (baM2) -- (baT);
			\draw[arrow] (baM2) -- (baC);
			\draw[arrow,dashed] (S) -- (baS);
			\draw[arrow,dashed] (S) -- (T);
			\draw[arrow,dashed] (baT) -- (S);
			\draw[arrow,dashed] (baB) -- (T);
			\draw[arrow,dashed] (baC) -- (S);
		\end{tikzpicture}}
	\end{center}
	\caption{An example of the loop operation $\oloop{}$.}
	\label{fig:loop}
\end{figure*}

We say an SPL graph $G=(V, E, S, T, B, C)$ is \emph{closed} if there are no incoming edges to the vertices $B$ and $C.$

\paragraph{SPLs as CFGs.} Given the above definitions of structured programs and SPL graphs, we have the following homomorphism which maps every structured program to its control-flow graph. Moreover, this homomorphism preserves closedness, i.e.~closed programs are mapped to closed graphs. A graph is an SPL graph if and only if it is the control-flow graph of a program~\cite{RegisterAllocation}. 

\begin{figure*}
	\hspace{-10pt}\begin{subfigure}[]{0.2\textwidth}
		\begin{lstlisting}
while x >= 1:
    if x >= y:
        x = x - y
        break
    else:
        y = y - x
        continue
		\end{lstlisting} 
	\end{subfigure}
\hfill
	 \begin{subfigure}[]{0.3\textwidth}\
     \scalebox{0.8}{
		\begin{tikzpicture}[scale=0.6]
			\node[rect] (eps1) {$\epsilon$};
			\node[rect] (break) [right=of eps1] {$\texttt{break}$};
			\node[rect] (eps2) [right=of break] {$\epsilon$};
			\node[rect] (continue) [right=of eps2] {$\texttt{continue}$};
			\node[rect] (sem1) [above=of eps1] {$;$};	
			\node[rect] (sem2) [above=of eps2] {$;$};	
			\node[rect] (if) [above=of sem1] {$\texttt{if}$};	
			\node[rect] (while) [above=of if] {$\texttt{while}$};	
			\draw[thick] (while) -- (if) ;
			\draw[thick] (if) -- (sem1) ;
			\draw[thick] (if) -- (sem2) ;
			\draw[thick] (sem1) -- (eps1) ;
			\draw[thick] (sem1) -- (break) ;
			\draw[thick] (sem2) -- (eps2) ;
			\draw[thick] (sem2) -- (continue) ;
		\end{tikzpicture}}
	\end{subfigure}
 \hspace{3cm}
\\
	\hspace{-10pt}\begin{subfigure}[]{0.2\textwidth}
    \scalebox{0.8}{
		\begin{tikzpicture}[scale=0.6]
			\node[rect] (eps1) {$A_\epsilon$};
			\node[rect] (break) [right=of eps1] {$A_\texttt{break}$};
			\node[rect] (eps2) [right=of break] {$A_\epsilon$};
			\node[rect] (continue) [right=of eps2] {$A_\texttt{continue}$};
			\node[rect] (sem1) [above=of eps1] {$\oseries{}{}$};	
			\node[rect] (sem2) [above=of eps2] {$\oseries{}{}$};	
			\node[rect] (if) [above=of sem1] {$\oparal{}{}$};	
			\node[rect] (while) [above=of if] {$\oloop{}$};	
			\draw[thick] (while) -- (if) ;
			\draw[thick] (if) -- (sem1) ;
			\draw[thick] (if) -- (sem2) ;
			\draw[thick] (sem1) -- (eps1) ;
			\draw[thick] (sem1) -- (break) ;
			\draw[thick] (sem2) -- (eps2) ;
			\draw[thick] (sem2) -- (continue) ;
		\end{tikzpicture}}
	\end{subfigure}
\hfill
	 \begin{subfigure}[]{0.3\textwidth}
     \scalebox{0.8}{
		\begin{tikzpicture}[scale=0.6,node distance=1cm and 2cm]
			
			\node[vertex,fill=yellow!20] (baM1)  {$M_1$};
			\node[vertex,fill=yellow!20] (baB)  [left=of baM1] {$B_1$};
			\node[vertex,fill=yellow!20] (baS) [right=of baM1] {$S_1$};
			\node[vertex,fill=yellow!20] (baM2) [right=of baS] {$M_2$};
			\node[vertex,fill=blue!20] (T) [above=of baM1] {$T$};
			
			\node[vertex] (S) [above=of baS] {$S$};
			\node[vertex,fill=yellow!20] (baC)  [above=of baM2] {$C_1$};

			\node[vertex,fill=red!20] (C) [below=of baS] {$C$};
			\node[vertex,fill=cyan!20] (B) [left=1cm of C] {$B$};
			
			\node[vertex,fill=yellow!20] (baT)  [right=1cm of C] {$T_1$};

			\draw[arrow] (S) -- (T) node [midway, above] {$x<1$};
			\draw[arrow] (S) -- (baS) node [midway, left] {$x\geq 1$};
			\draw[arrow] (baS) -- (baM1) node [midway, above] {$x \geq y$};
			\draw[arrow] (baS) -- (baM2) node [midway, above] {$x < y$};
			\draw[arrow] (baM1) -- (baB) node [midway, above] {$x \gets x - y$};
			\draw[arrow] (baM2) -- (baC) node [midway, right] {$y \gets y - x$};
			\draw[arrow] (baB) -- (T) node [midway, left] {$\texttt{break}$};
			\draw[arrow] (baC) -- (S) node [midway, above] {$\texttt{continue}$};
			\draw[arrow] (baT) -- (S);
		\end{tikzpicture}}
	\end{subfigure}
 \hspace{4.8cm}
 
	\caption{SPL decomposition example}
	\label{fig:decompo}
\end{figure*}
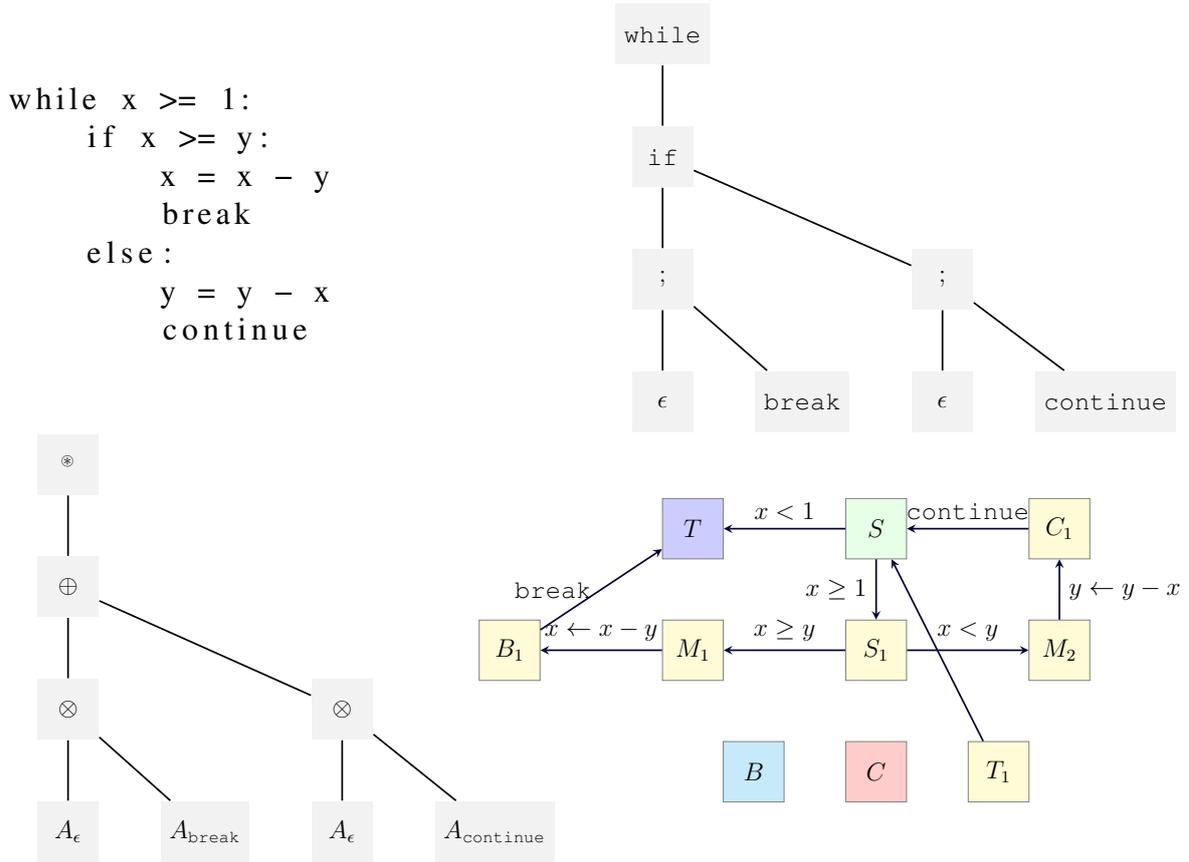

\paragraph{SPL Decomposition.} Given a closed program $P,$ we can first parse it based on the grammar in~\eqref{gram:prog} to generate a parse tree. Subsequently, by applying our homomorphism above to this parse tree, we can derive a parse tree according to~\eqref{gram:spl} for its control-flow graph. We use the term \emph{SPL decomposition} to refer to the parse tree of the CFG according to~\eqref{gram:spl}. It is easy to verify that this process takes linear time. See Figure~\ref{fig:decompo} as an example, where the program $P$ is at top left, its parse tree is at top right, the corresponding parse tree of $G = \cfg{P},$ aka the grammatical decomposition of $G$ is at bottom left and the graph $G = \cfg{P}$ is the bottom right one. The edges of the graph are labelled according to the commands/conditions of the program.

\newpage

\chapter{Register Allocation}
\label{chp_background}

Register allocation \cite{BouchezmRegister2006} is a vital component of compiler optimization, involving the assignment of a finite number of processor registers to a potentially extensive set of variables utilized within a program. In contemporary programming languages, programmers can define numerous variables as needed. However, the compiler must determine which variables to allocate in the processor's registers and which to store in the main memory (RAM). Accessing registers is considerably faster than accessing main memory, but the number of available registers is often quite limited. Given a program \( P \) and an integer \( r \), the register allocation problem seeks to ascertain whether it is feasible to assign the variables in \( P \) to \( r \) registers without necessitating access to main memory (no spilling). More specifically, if two variables \( x \) and \( y \) may be alive simultaneously, they interfere with each other and must be assigned to different registers. A variable is considered alive at a particular point in the program if it has been assigned a value that may be used in the future. 

Now we considered the problem of minimum-cost register allocation as formalized in~\cite{DBLP:conf/cc/Krause13}. A cost is assigned to each allocation of variables to registers, which is supposed to model the time wasted on spills or rematerialization. We note that this is a more general formulation of the problem than those of~\cite{DBLP:journals/iandc/Thorup98,conradof2023bounded}, which only focus on deciding whether it is possible to avoid spilling altogether and obtain a cost of zero.

Suppose we are given a program $P$ with control-flow graph $G=\cfg{P} = (V, E, S, T, B, C).$ Let $[r] = \{0, 1, \ldots, r-1\}$ be the set of available registers and $\vars$ the set of our program variables. Every variable $v \in \vars$ has a lifetime $\lt{v}$ which is a connected subgraph of $G.$ See~\cite{poletto1999linear} for a more detailed treatment of lifetimes. Since lifetimes can be computed by a simple data-flow analysis, we assume without loss of generality that they are given as inputs to our algorithm. For a vertex $v$ or edge $e$ of $G,$ we denote the set of variables that are alive at this vertex/edge by $L(v)$ or $L(e).$ An \emph{assignment} is a function $f: \vars \rightarrow [r] \cup \{\perp\}$ which maps each variable either to a register or to $\perp.$ The latter models the variable being spilled. An assignment is valid if it does not map two variables with intersecting lifetimes to the same register. We denote the set of all valid assignments by $F.$

The interference graph of our program $P$ is a graph $\mathbb{I} = (\vars, E_{\mathbb I})$ in which there is one vertex for each program variable and there is an edge $\{u, v\}$ if the variables $u$ and $v$ can be alive at the same time, i.e.~$\lt{u} \cap \lt{v} \neq \emptyset.$ Any valid assignment $f$ is a valid coloring of a subset of vertices of $\mathbb{I}$ with colors in $[r].$ This correspondence between register allocation and graph coloring is well-known and due to Chaitin~\cite{DBLP:conf/pldi/Chaitin82}. We note that for every vertex $v,$ the set $L(v)$ of variables alive at $v$ forms a clique in $\mathbb I.$

\begin{figure*}
	\begin{subfigure}{0.33\textwidth}
		\begin{lstlisting}[mathescape,numbers=none]
while $\varphi_1$ do
	$a \gets b+c$;
	$d \gets -a$;
	$e \gets d+f$;
	if $\varphi_2$ $\texttt{then}$
		$f \gets 2 \cdot e$;
	else
		$b \gets d + e$;
		$e \gets e - 1$;
	fi
	$b \gets f + c$;
od
		\end{lstlisting} 
	\end{subfigure}
	\begin{subfigure}{0.33\textwidth}
		\begin{center}
        \scalebox{0.8}{
		\begin{tikzpicture}[scale=0.6]
			\node[vertex] (s)  {$S$};
			\node[vertexla] (ls) [above=0cm of s]  {$\{b, c, f\}$};
			\node[vertex,fill=blue!20] (t) [right=of s] {$T$};
			\node[vertexla] (lt) [above=0cm of t]  {$\emptyset$};
			\node[vertex,fill=cyan!20] (b) [right=of t]  {$B$};
			\node[vertexla] (lb) [above=0cm of b]  {$\emptyset$};
			\node[vertex,fill=red!20] (c) [right=of b]  {$C$};
			\node[vertexla] (lc) [above=0cm of c]  {$\emptyset$};
			\node[vertex,fill=yellow!20] (c1) [below=of t]  {$C_1$};
			\node[vertexla] (lc1) [below=0cm of c1]  {$\{b, c, f\}$};
			\node[vertex,fill=yellow!20] (b1) [right=of c1]  {$B_1$};
			\node[vertexla] (lb1) [below=0cm of b1]  {$\emptyset$};
			\node[vertex,fill=yellow!20] (1) [below=of s]  { };
			\node[vertexla] (l1) [left=0cm of 1]  {$\{b, c, f\}$};
			\node[vertex,fill=yellow!20] (2) [below=of 1]  { };
			\node[vertexla] (l2) [left=0cm of 2]  {$\{a, c, f\}$};
			\node[vertex,fill=yellow!20] (3) [below=of 2]  { };
			\node[vertexla] (l3) [left=0cm of 3]  {$\{c, d, f\}$};
			\node[vertex,fill=yellow!20] (if) [below=of 3]  { };
			\node[vertexla] (lif) [left=0cm of if]  {$\{c, d, e, f\}$};
			\node[vertex,fill=yellow!20] (then) [right=3cm of if]  { };
			\node[vertexla] (lthen) [right=0cm of then]  {$\{c, e, f\}$};
			\node[vertex,fill=yellow!20] (fi) [below=of if] {};
			\node[vertexla] (lfi) [left=0cm of fi]  {$\{c, f\}$};
			\node[vertex,fill=yellow!20] (5) [below=of fi] {};
			\node[vertexla] (l5) [left=0cm of 5]  {$\{b, c, f\}$};
            
			\draw[arrow] (1) -- (2) node [midway, left] {$a \gets b+c$};
			\draw[arrow] (2) -- (3) node [midway, left] {$d \gets -a$};
			\draw[arrow] (3) -- (if) node [midway, left] {$e \gets d+f$};
			\draw[arrow] (if) -- (then) node [midway, above] {$\neg\varphi_2, b \gets d+e$};
			\draw[arrow] (if) -- (fi) node [midway, left] {$\varphi_2, f \gets 2 \cdot e$};
			\draw[arrow] (then) -- (fi) node [midway, below right] {$e \gets e-1$};
			\draw[arrow] (fi) -- (5) node [midway, right] {$b \gets f+c$};
			\draw[arrow] (s) -- (1) node [midway, left] {$\varphi_1$};
			\draw[arrow] (s) -- (t) node [midway, above] {$\neg \varphi_1$};
			\draw [arrow] (5) to [out=150,in=210] (s);
			%\draw[arrow, bend left=90] (5) -- (s);
			\draw[arrow] (b1) -- (t);
			\draw[arrow] (c1) -- (s);
		\end{tikzpicture}}
	\end{center}
	\end{subfigure}

\vspace{1em}
	
		\begin{subfigure}{0.25\textwidth}
        \scalebox{0.8}{
			\begin{tikzpicture}[scale=0.6]
				\node[vertex] (a)  {$a$};
				\node[vertex] (b) [below=of a] {$b$};
				\node[vertex] (c) [right=of a] {$c$};
				\node[vertex] (f) [right=of b] {$f$};
				\node[vertex] (e) [right=of c] {$e$};
				\node[vertex] (d) [right=of f] {$d$};
				\draw[thick] (b) -- (c);
				\draw[thick] (b) -- (f);
				\draw[thick] (f) -- (c);
				\draw[thick] (a) -- (c);
				\draw[thick] (a) -- (f);
				\draw[thick] (d) -- (c);
				\draw[thick] (d) -- (f);
				\draw[thick] (e) -- (c);
				\draw[thick] (e) -- (d);
				\draw[thick] (e) -- (f);
			\end{tikzpicture}}
	\end{subfigure}
    \hfill
\begin{subfigure}{0.25\textwidth}
\scalebox{0.8}{
	\begin{tikzpicture}[scale=0.6]
		\node[vertex, fill=blue!40] (a)  {$a$};
		\node[vertex, fill=green!40] (b) [below=of a] {$b$};
		\node[vertex, fill=red!40] (c) [right=of a] {$c$};
		\node[vertex, fill=yellow!40] (f) [right=of b] {$f$};
		\node[vertex, fill=green!40] (e) [right=of c] {$e$};
		\node[vertex, fill=blue!40] (d) [right=of f] {$d$};
		\draw[thick] (b) -- (c);
		\draw[thick] (b) -- (f);
		\draw[thick] (f) -- (c);
		\draw[thick] (a) -- (c);
		\draw[thick] (a) -- (f);
		\draw[thick] (d) -- (c);
		\draw[thick] (d) -- (f);
		\draw[thick] (e) -- (c);
		\draw[thick] (e) -- (d);
		\draw[thick] (e) -- (f);
	\end{tikzpicture}}
\end{subfigure}
\hfill
\begin{subfigure}{0.25\textwidth}
\scalebox{0.8}{
	\begin{tikzpicture}[scale=0.6]
		\node[vertex] (a)  {$a$};
		\node[vertex, fill=green!40] (b) [below=of a] {$b$};
		\node[vertex, fill=red!40] (c) [right=of a] {$c$};
		\node[vertex] (f) [right=of b] {$f$};
		\node[vertex, fill=green!40] (e) [right=of c] {$e$};
		\node[vertex, fill=blue!40] (d) [right=of f] {$d$};
		\draw[thick] (b) -- (c);
		\draw[thick] (b) -- (f);
		\draw[thick] (f) -- (c);
		\draw[thick] (a) -- (c);
		\draw[thick] (a) -- (f);
		\draw[thick] (d) -- (c);
		\draw[thick] (d) -- (f);
		\draw[thick] (e) -- (c);
		\draw[thick] (e) -- (d);
		\draw[thick] (e) -- (f);
	\end{tikzpicture}}
\end{subfigure}
\caption{Register Allocation Example}
\label{fig:regaloc}
\end{figure*}
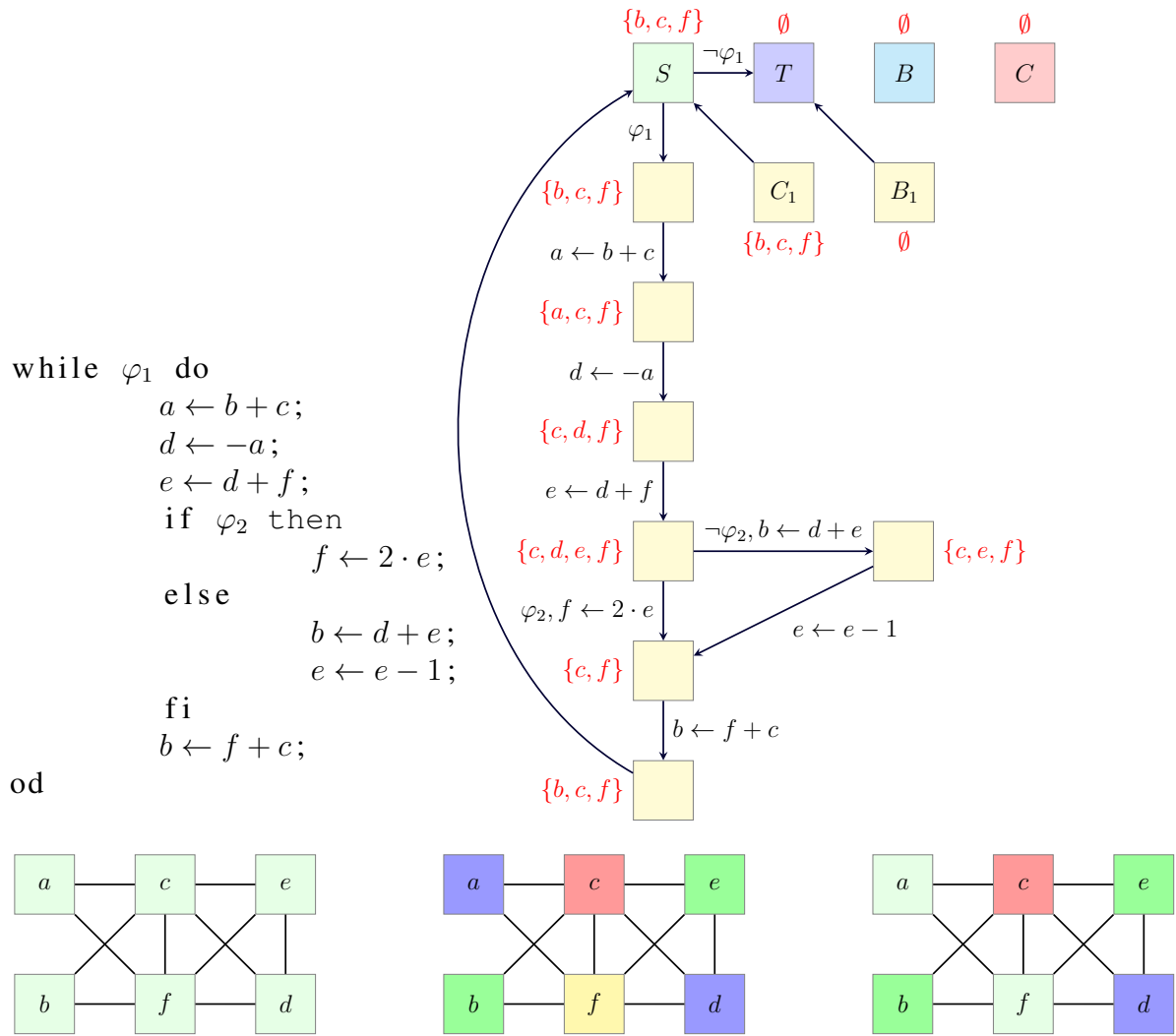

We provide an example taken from~\cite{example}. Figure~\ref{fig:regaloc} shows a program $P$ and its control-flow graph $G = \cfg{P},$ including live variables at each vertex, and the interference graph $\mathbb{I}.$ The program $P$ is at top left, its control-flow graph $G = \cfg{P}$ is the top right one in which every vertex is labeled by its set of live variables in red. The interference graph $\mathbb I$ is at bottom left, while a coloring of all vertices of $\mathbb{I}$ with 4 colors corresponding to allocating all variables to 4 registers is at bottom center, and a coloring of a subset of vertices of $\mathbb{I}$ with 3 colors corresponding to spilling the variables $a$ and $f$ is at bottom right. Our goal is to color a subset of vertices of $\mathbb{I}$ with $r$ colors, where $r$ is the number of available registers. A complete coloring with $4$ colors is shown in the figure. This avoids any spilling. We also show a partial coloring with $3$ colors and some spilling.

Now let's try to consider this problem as PCSP. As all spills and rematerialization have happened during the edge, the cost function is $c: E \times A \times A \rightarrow [0, \infty).$ For an edge $e \in E$ of the control-flow graph, which corresponds to one command of the program, $c(e, a_1, a_2)$ is the cost of running this command when the alive registers are allocated as $a_1$ before entering $e$ and as $a_2$ when leaving $e.$ We assume that the cost at $e$ only depends on the allocation decisions for variables that are alive at $e.$, which is union of $a_1$ and $a_2$, hence if they have any contradictions like assign the same variable to different registers, we can directly set the cost as $\infty$. Following~\cite{DBLP:conf/cc/Krause13}, we further assume constant-time oracle access to evaluations of $c.$ In practice, $c$ is often obtained by profiling. Different optimization goals, such as total runtime or code size, may be modeled by choosing a suitable function $c.$

\paragraph{Our Algorithm.}
We now show how to perform dynamic programming on the grammatical decomposition of our control-flow graph $G$ to find an optimal register allocation. Our algorithm is quite simple and elegant. We process our grammatical decomposition in a bottom-up fashion, and for every subgraph $H = (V_H, E_H, S_H, T_H, B_H, C_H)$ appearing in the grammatical decomposition, define the following dynamic programming variables:
$$
\begin{array}{rl}
\opt{H, f'} = & \text{The minimum total cost $\sum_{e \in E_H} c(e, f)$}\\
& \text{of an assignment $f$ over $H$ such that} \\ 
& f_{|L(S_H) \cup L(T_H) \cup L(B_H) \cup L(C_H)}=f'.
\end{array}$$
Intuitively, for every possible assignment $f'$ of the variables that are alive at any of the distinguished vertices $(S_H, T_H, B_H, C_H),$ we are asking for the minimum total cost of an assignment $f$ over all variables that agrees with $f'$ and extends it. After we compute our $\opt{\cdot, \cdot}$ values, the final answer of the algorithm, i.e.~the minimum cost of a register allocation, is simply $\min_{f} \opt{G, f}.$

We now show how to compositionally compute $OPT[H, f']$ for any $SLP$ graph $H$ assuming that we have already computed $OPT[\cdot, \cdot]$ values for the $SLP$ subgraphs of $H.$ This is done by casework:

\begin{itemize}

	\item \emph{Atomic Graphs:} 	
	If $H \in \{A_\epsilon, A_{\texttt{break}}, A_{\texttt{continue}}\},$ then $H$ does not have any vertices other than the distinguished vertices $(S_H, T_H, B_H, C_H).$ Thus, all variables that are alive at any point in $H$ are also alive at one of the distinguished vertices, and we simply set $\opt{H, f'} = c(e, f')$ for every partial allocation $f'.$ Here, $e$ is the unique edge in $H.$
\end{itemize}
	
	\textbf{Compatible Assignments} We say two partial assignments $f_1: \vars_1 \rightarrow [r] \cup \{\perp\}$ and $f_2: \vars_2 \rightarrow [r] \cup \{\perp\}$ are compatible and write $f_1 \leftrightarrows f_2$ if $\forall v \in \vars_1 \cap \vars_2 \quad f_1(v) = f_2(v).$
	Informally, $f_1$ and $f_2$ never make conflicting decisions on any variable $v$ but we have no restrictions on variables that are decided only by $f_1$ or only by $f_2.$ In other words, $f_1$ and $f_2$ can be combined in the same total assignment.

\begin{itemize}

	\item \emph{Series Operation:} If $H = \oseries{H_1}{H_2},$ then we have
	
	 $$\opt{\oseries{H_1}{H_2}, f'} =$$ $$\min_{\begin{matrix}f' \leftrightarrows f_1'\\ f' \leftrightarrows f_2'\\ f_1' \leftrightarrows f_2'\end{matrix}} \left(\opt{H_1, f_1'} + \opt{H_2, f_2'} - \sum_{e \in E_{H_1} \cap E_{H_2}} c(e, f_1') \right).$$

	The correctness of this calculation is an immediate corollary of the definition of our series operation. By construction, we have $L(B_{H_1}) = L(B_{H_2}) = L(B_H)$ and $L(C_{H_1}) = L(C_{H_2}) = L(C_H)$ and also $L(T_{H_1}) = L(S_{H_2}).$ Moreover, every edge of $H_1$ and $H_2$ is preserved in $\oseries{H_1}{H_2}.$ Thus, the total cost is simply the sum of costs in the two components. We should also be careful not to double-count the cost of edges that appear in both $H_1$ and $H_2$. Thus, we subtract these. Of course, the partial assignments $f', f_1'$ and $f_2'$ should be pairwise compatible.
	
	\item \emph{Parallel Operation:} This case is handled exactly as in the series case:
	$$
	\displaystyle \opt{\oparal{H_1}{H_2}, f'} =$$ $$\displaystyle \min_{\begin{matrix}f' \leftrightarrows f_1'\\ f' \leftrightarrows f_2'\\ f_1' \leftrightarrows f_2'\end{matrix}} \left(\opt{H_1, f_1'} + \opt{H_2, f_2'} - \sum_{e \in E_{H_1} \cap E_{H_2}} c(e, f_1') \right).
	$$
	This is because our parallel operation also preserves all the edges in $H_1$ and $H_2.$ Note that we might have edges that appear in both $H_1$ and $H_2,$ e.g.~we might have both $(S_{H_1}, B_{H_1})$ and $(S_{H_2}, B_{H_2})$ which are the same as the edge $(S_H, B_H).$ Thus, the total cost is the sum of the costs in the components $H_1$ and $H_2$ minus the cost of their common edges. As before, we should also ensure that the partial assignments are all compatible.
	
	\item \emph{Loop Operation:} Suppose $H = \oloop{H_1}.$ In this case, by our construction, $H$ has the same vertices and edges as $H_1$ except for the introduction of the four new distinguished vertices $(S_H, T_H, B_H, C_H)$ and five new edges $e_1 = (S_H, S_{H_1}), e_2 = (S_H, T_H), e_3 = (T_{H_1}, S_H), e_4 = (C_{H_1}, S_H)$ and $e_5 = (B_{H_1}, T_H).$ Thus, our total cost is simply the total cost in $H_1$ plus the cost incurred at these new edges. Therefore, we have:
	$$
	\opt{\oloop{H_1}, f'} = \min_{f_1' \leftrightarrows f'} \left( \opt{H_1, f_1'} + \sum_{i=1}^5 c(e_i, f' \cup f_1') \right).
	$$
\end{itemize}
This concludes our algorithm, which computes the cost of an optimal assignment $f.$ As is standard in dynamic programming approaches, $f$ itself can be obtained by retracing the steps of the algorithm and remembering the choices that led to the minimum values at every step.

\paragraph{\textbf{Theorem 4.1}}
	Given a program $P$ with variables $\vars$ and control-flow graph $G=\cfg{P},$ the number $r$ of available registers and a cost function $c(\cdot, \cdot)$ as input, our algorithm above finds an optimal allocation of registers, i.e.~an optimal assignment function $f,$ in time $O(|G| \cdot |\vars|^{5 \cdot r}).$\\
\textbf{Proof.}
	Correctness was argued above. We do casework for runtime analysis:
	%\begin{itemize}
		%\item
		At atomic graphs, we are considering partial assignments $f'$ over variables that are alive at any of the four distinguished vertices. Let $a$ be one of these distinguished vertices. The set $L(a)$ of alive variables at $a$ forms a clique in the interference graph $\mathbb I.$ Thus, any valid $f'$ can assign $f'(v) \neq \perp$ to at most $r$ variables $v$ in $L(a).$ Moreover, no two variables can be assigned to the same register. Given that $|L(a)| \leq |\vars|,$ the total number of possible assignments for variables in $L(a)$ is at most $$\binom{|\vars|}{r} \cdot r! + \binom{|\vars|}{r-1} \cdot (r-1)! + \dots + \binom{|\vars|}{0} \cdot 0! \in O(r \cdot |\vars|^r).$$ Thus, the total number of $f'$ functions is at most $O(r^4 \cdot |\vars|^{4 \cdot r})$ given that we have four distinguished vertices. Our algorithm spends a constant amount of time for each $f',$ simply querying the cost of a single edge.
		%\item
		
		When $H = \oseries{H_1}{H_2},$ we note that we have $B_H = B_{H_1} = B_{H_2}$ and $C_H = C_{H_1} = C_{H_2}.$ Similarly, we have $T_{H_1} = S_{H_2}.$ Thus, $f', f_1'$ and $f_2'$ need to jointly choose a register assignment for the variables that are alive at one of five vertices: $S_{H_1}, T_{H_1}, T_{H_2}, B$ and $C.$ An argument similar to the previous case shows that there are $O(r^5 \cdot |\vars|^{5 \cdot r})$ such assignments. We also note that $E_{H_1} \cap E_{H_2}$ has $O(1)$ many edges since any such edge must be connecting two distinguished vertices, and we have only four such vertices. Thus, the total runtime here is also $O(r^5 \cdot |\vars|^{5 \cdot r}).$
		
		%\item
		
		When $H = \oparal{H_1}{H_2},$ a similar argument applies. In this case, we have $S_H = S_{H_1} = S_{H_2}, T_H = T_{H_1} = T_{H_2}, B_H = B_{H_1} = B_{H_2}$ and $C_H = C_{H_1} = C_{H_2}.$ Thus, we need to look at assignments for live variables at only four different vertices, and our runtime is $O(r^4 \cdot |\vars|^{4 \cdot r}).$
		
		%\item
		
		Finally, when $H = \oloop{H_1},$ we are introducing four new distinguished vertices. So, it seems that we have to consider the live variables at eight vertices in total, the distinguished vertices of both $H$ and $H_1.$ However, note that $B_{H_1}$ has only one outgoing edge in our control-flow graph $G$ which goes to $T_H.$ Thus, we have $L(B_{H_1}) \subseteq L(T_H).$ For similar reasons, $L(T_{H_1}) \subseteq L(S_H)$ and $L(C_{H_1}) \subseteq L(S_H).$ Therefore, we only need to consider the program variables that are alive at one of the five vertices $S_H, T_H, B_H, C_H$ and $S_{H_1}.$ An argument similar to the previous cases shows that our runtime is $O(r^5 \cdot |\vars|^{5 \cdot r}).$
%	\end{itemize}

	Finally, our algorithm has to process the grammatical decomposition in a bottom-up manner and compute the $\opt{\cdot, \cdot}$ values at every node. We have $O(|G|)$ nodes. Thus, the total worst-case runtime is $O(|G| \cdot r^5 \cdot |\vars|^{5 \cdot r}).$ Following~\cite{DBLP:conf/cc/Krause13} and other works on minimum-cost register allocation, we assume that $r$ is a constant. Thus, our runtime is $O(|G| \cdot |\vars|^{5 \cdot r}).$

\newpage

\chapter{Lifetime-optimal Speculative Partial Redundancy Elimination}
\label{chp_background}

Redundancy elimination (RE), i.e.~avoiding repeated and unnecessary computations of the same expression, has been a goal of optimizing compilers since their early days. Put simply, if the same expression $e$ is used in several different locations in a program, it might be beneficial to compute $e$ once, store it in a temporary variable, and then use it whenever the program reaches any of the locations that need $e.$ Here is a simple example, considering the following program.
\begin{lstlisting}
int f(int a, int b){
    if(a+b>3){
        return a+b-3;
    }
    else {
        return a+b;
    }
}
\end{lstlisting}
       
\begin{figure*}[h!]
    \centering
        \includegraphics[scale= 0.7]{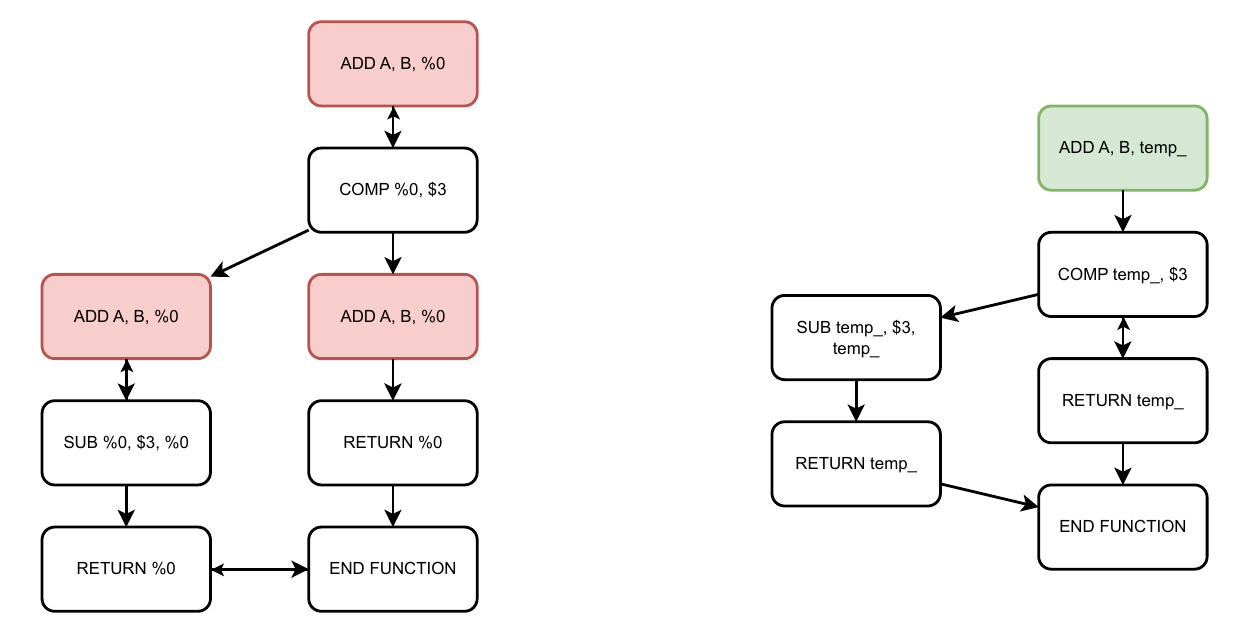}       
  
    \caption{Intermediate representation (IR) before and after optimization.}
    \label{fig:RE}
\end{figure*}

In this case, we calculate $a+b$ three times. Its intermediate representation before and after the RE optimization would look like Figure~\ref{fig:RE}. Notably, the code size is smaller after the optimization.

One of the first formalizations of this problem was provided in 1970 as Global Common Subexpression Elimination (GCSE)~\cite{cocke1970GCSE}. Later approaches considered removing redundancies that appear only in a subset of paths of the control-flow graph, leading to Partial Redundancy Elimination (PRE)~\cite{morel1979partial}. An enhancement to PRE, introduced by Lazy Code-Motion (LCM)~\cite{knoop1992lazycodemotion}, focuses on achieving 
lifetime optimality by minimizing the lifetimes of the temporary variables it introduces. This is also helpful for reducing register pressure. Another classical improvement is that of Speculative PRE (SPRE)~\cite{cai2003SPRE,gupta1998profile}, which selects the path for adding computations based on profiling 
information with the goal of maximizing the benefits of PRE. Putting the ideas of LCM and SPRE together leads to Lifetime-Optimal SPRE (LOSPRE), which is currently the most expressive approach to redundancy elimination and subsumes all other methods mentioned above.

Now with the CFG $G=\{V,E\}$, we can define this problem as following

\begin{itemize}
    \item \emph{Use set} Consider an expression $e.$ We define the use set $U$ of $e$ as the set of all nodes of the CFG in which the expression $e$ is computed.

    \item \emph{Life set} Our goal is to precompute the expression $e$ at a few points, save the result in a temporary variable \texttt{temp}, and then use   \texttt{temp} in place of $e$ in every node of $U.$ We denote the lifetime of the variable $\texttt{temp}$ by $L$ and call it our life set. 

    \item \emph{Invalidating set}We say a node $v$ of the CFG invalidates $e$ if the statement at $v$ changes the value of $e.$ For example, if $e = \texttt{a+b},$ then the statement $\texttt{a = 0}$ invalidates $e.$ We denote the set of all invalidating nodes by $I.$ These nodes play a crucial role in LOSPRE since they force us to update the value saved in $\texttt{temp}$ by recomputing $e.$ We assume that the entry and exit nodes are invalidating since LOSPRE is an intraprocedural analysis that has no information about the program's execution before or after the current function.

    \item \emph{Calculating set} Given the sets $U, L$ and $I$ above, we have to make sure the value of our temporary variable $\texttt{temp}$ is correct at every node in $U \cup L.$ Thus, for every edge $(x, y) \in E$ of the CFG where $x \not \in L$ and $y \in U \cup L,$ we have to insert a computation $\texttt{temp} = e$ between $x$ and $y.$ Similarly, if $x \in I,$ then the value stored at $\texttt{temp}$ becomes invalid after the execution of $x,$ requiring us to inject the same computation between $x$ and $y.$ Formally, the computation $\texttt{temp} = e$ has to be injected into the following set of edges of the CFG:
$$
C(U, L, I) = \{ (x, y) \in E ~\vert~ x \not\in L \setminus I ~\land~ y \in U \cup L \}.
$$

\end{itemize}

\begin{figure}
	\begin{center}
		\begin{tikzpicture}[scale=0.6]
			\node[redvertex] (1) {$1$};
			\node[grayvertex] (2)  [right =of 1] {$2$};
			\node[whitevertex] (3)  [right =of 2] {$3$};
			\node[grayvertex] (4)  [right =of 3] {$4$};
			\node[grayvertex] (5)  [below =of 4] {$5$};
			\node[redvertex] (6)  [right =of 4] {$6$};
			\node[grayvertex] (7)  [right =of 6] {$7$};
			\node[redvertex] (8)  [right =of 7] {$8$};
			\draw[arrow] (1) -- (2);
			\draw[arrow] (2) -- (3);
			\draw[arrow] (3) -- (4);
			\draw[arrow] (3) -- (5);
			\draw[arrow] (5) -- (6);
			\draw[arrow] (4) -- (6);
			\draw[arrow] (6) -- (7);
			\draw[arrow] (7) -- (8);
		\end{tikzpicture}
	\end{center}
	
	\begin{center}
		\begin{tikzpicture}[scale=0.6]
			\node[redvertex] (1) {$1$};
			\node[bluevertex] (2)  [right =of 1] {$2$};
			\node[bluevertex] (3)  [right =of 2] {$3$};
			\node[grayvertex] (4)  [right =of 3] {$4$};
			\node[grayvertex] (5)  [below =of 4] {$5$};
			\node[redvertex] (6)  [right =of 4] {$6$};
			\node[grayvertex] (7)  [right =of 6] {$7$};
			\node[redvertex] (8)  [right =of 7] {$8$};
			\draw[arrow,blue] (1) -- (2);
			\draw[arrow] (2) -- (3);
			\draw[arrow] (3) -- (4);
			\draw[arrow] (3) -- (5);
			\draw[arrow] (5) -- (6);
			\draw[arrow] (4) -- (6);
			\draw[arrow,blue] (6) -- (7);
			\draw[arrow] (7) -- (8);
		\end{tikzpicture}
	\end{center}
	
	\begin{center}
		\begin{tikzpicture}[scale=0.6]
			\node[redvertex] (1) {$1$};
			\node[grayvertex] (2)  [right =of 1] {$2$};
			\node[bluevertex] (3)  [right =of 2] {$3$};
			\node[grayvertex] (4)  [right =of 3] {$4$};
			\node[grayvertex] (5)  [below =of 4] {$5$};
			\node[redvertex] (6)  [right =of 4] {$6$};
			\node[grayvertex] (7)  [right =of 6] {$7$};
			\node[redvertex] (8)  [right =of 7] {$8$};
			\draw[arrow,blue] (1) -- (2);
			\draw[arrow, blue] (2) -- (3);
			\draw[arrow] (3) -- (4);
			\draw[arrow] (3) -- (5);
			\draw[arrow] (5) -- (6);
			\draw[arrow] (4) -- (6);
			\draw[arrow,blue] (6) -- (7);
			\draw[arrow] (7) -- (8);
		\end{tikzpicture}
	\end{center}
	\caption{An example of LOSPRE}
	\label{fig:ex}
\end{figure}
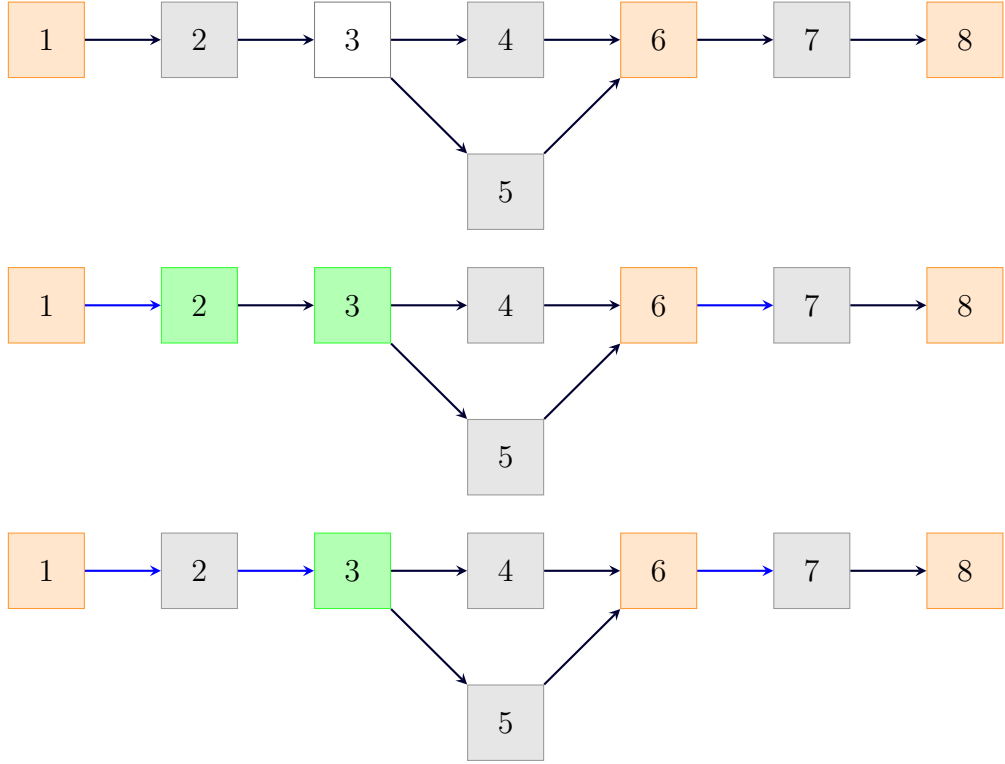

Figure~\ref{fig:ex} shows an example of LOSPRE.  The top part of the figure is a CFG in which the use set of an expression $e$ is shown in gray. We need the value of $e$ at the vertices $U = \{2, 4, 5, 7\}.$ The invalidating set is shown in orange, i.e.~the vertices in $I = \{1, 6, 8\}$ invalidate $e.$ The middle and bottom parts each show one possible optimization. We show the lifetime of our temporary variable in green. 

In the middle part, the temporary variable is alive at $\{2, 3\}.$ Thus, the computation $\texttt{temp} = e$ has to be injected into the edge $(1, 2).$ We can then use $\texttt{temp}$ instead of $e$ in locations $2, 4$ and $5.$ However, we need to recompute $e$ in the edge $(6, 7).$ In this case our computation set is $\{(1, 2), (6, 7)\}.$ The edges in the computation set are shown in blue.

In the bottom part, the temporary variable is alive only at position $3.$ Thus, we first compute $e$ when passing through $(1, 2)$ so that we have its value at $2.$ We then recompute $e$ when going through $(2, 3)$ and save it at a temporary variable $\texttt{temp}.$ This temporary variable is then used in place of $e$ in $4$ and $5.$ This example shows a tradeoff in which fewer repetitions of the computation lead to a longer lifetime for the temporary variable, which increases register pressure and is undesirable for register allocation.

This time, there are two types of costs associated with the process above: (i)~injecting calculations into the edges in $C(U, L, I)$ and (ii)~keeping an extra variable $\texttt{temp}$ at every node in $L.$ These costs are dependent on the goals pursued by the compiler. For example, a compiler aiming to minimize code size will focus on (i). On the other hand, if our goal is to ease register pressure, we would want to minimize (ii). LOSPRE is an expressive framework in which these costs are modeled by two functions
$$
c: E \rightarrow K
$$ 
and 
$$
l: V \rightarrow K.
$$
where $K$ is a totally-ordered set with an addition operator, $c$ is a function that maps each edge to the cost of adding a computation of $e$ in that edge and $l$ is similarly a function that maps each vertex of the CFG to the cost of keeping the temporary variable $\texttt{temp}$ alive at that vertex.

Based on the discussion above, we are now ready to define our main problem.

Given a CFG $G=(V,E)$, a use set $U,$ an invalidating set $I$ and two cost functions $c: E \rightarrow K$ and $l: V \rightarrow K,$ the LOSPRE problem is to find a life set $L$ that minimizes the total cost $$\cost(G, U, I, L, c, l) = \sum_{e\in C(U,L,I)}c(e)+\sum_{v\in L} l(v).$$

\paragraph{Our Algorithm.}
In this section, we present a linear-time algorithm for LOSPRE using SPL decompositions. The input to our algorithm consists of a closed program $P,$ its control-flow graph $G=(V, E),$ a use set $U \subseteq V,$ an invalidating set $I \subseteq V$ and two cost functions $c: E \rightarrow K$ and $l: V \rightarrow K.$ Our goal is to find a life set $L \subseteq V$ that minimizes
$$\cost(G, U, I, L, c, l) = \sum_{e\in C(U,L,I)}c(e)+\sum_{v\in L} l(v).$$

\textbf{Step 1 (Initialization)} Our algorithm computes an SPL decomposition of $G = \cfg{P}$ by first parsing $P$ and then applying the homomorphism of the previous section. 

\textbf{Step 2 (Dynamic Programming)} Our algorithm proceeds with a bottom-up dynamic programming on the SPL decomposition. Note that each node $u$ of the SPL decomposition corresponds to an SPL subgraph $G_u = (V_u, E_u, S_u, T_u, B_u, C_u)$ of $G$ which is either an atomic SPL graph (when $u$ is a leaf) or obtained by applying one of the SPL operations to the graphs corresponding to the children of $u.$ See Figure~\ref{fig:decompo}. Let $\varGamma_u = \{S_u, T_u, B_u, C_u\}$ be the set of special vertices of $G_u.$ For every $X \subseteq \varGamma_u,$ we define a dynamic programming variable $\DP[u, X].$ Our goal is to compute this dynamic programming value such that
$$
\DP[u, X] = \min_{L \subseteq V_u ~\land~ L \cap \varGamma_u = X} \cost(G_u, U, I, L, c, l).
$$
Intuitively, we are considering a subproblem of the original LOSPRE in which the graph is limited to $G_u.$ Moreover, we only consider those solutions (life sets) $L$ for which $L \cap \varGamma_u = X.$ The value in $\DP[u, X]$ should then give us the minimum cost among all such solutions. Below, we present how our algorithm computes $\DP[u, X]$ for every vertex $u$ of the decomposition based on the $\DP[\cdot, \cdot]$ values at its children:
\begin{itemize}
	\item \emph{Atomic Nodes}: If $G_u$ is an atomic SPL graph, then the only vertices in $G_u$ are the four special vertices. Therefore, we must have $L=X.$ Our algorithm computes each $\DP[u, X]$ as:
	 $$\DP[u, X] = \cost(G_u, U, I, X, c, l) = \sum_{e\in C(U,X,I) \cap G_u} c(e)+\sum_{v\in X} l(v).$$
	 
	 \item \emph{Series Nodes}: Suppose $G_u = \oseries{G_v}{G_w}$ where $v$ and $w$ are the children of $u$ in the SPL decomposition. Let $X \subseteq \varGamma_u$ and $X_v \subseteq \varGamma_v$ be subsets of special vertices of $G_u$ and $G_v,$ respectively. We say that $X$ and $X_v$ are \emph{compatible} and write $X \compatible X_v$
	 if the following conditions are satisfied:
	 \begin{itemize}
	 \item 	$S_v \in X_v \Leftrightarrow S_u \in X;$
	 \item	$B_v \in X_v \Leftrightarrow B_u \in X;$
	 \item	$C_v \in X_v \Leftrightarrow C_u \in X.$
	 \end{itemize}
 	Intuitively, compatibility means that the subsets $X$ and $X_v$ make the same decisions about including vertices in the life set $L.$ Since $S_u = S_v,$ they should either both include it or both exclude it. Similarly, $B_u$ is obtained by merging $B_v$ and $B_w$. Therefore, the decisions made for $B_u$ and $B_v$ must match. The same applies to $C_u$, which is a merger of $C_v$ and $C_w.$ 
 	
 	Now consider $X_w \subseteq \Gamma_w.$ We say that $X_w$ and $X$ are compatible and write $X \compatible X_w$ if the following conditions are satisfied:
 	\begin{itemize}
 		\item 	$T_w \in X_w \Leftrightarrow T_u \in X;$
 		\item	$B_w \in X_w \Leftrightarrow B_u \in X;$
 		\item	$C_w \in X_w \Leftrightarrow C_u \in X.$
 	\end{itemize}
 	The intuition is the same as the previous case, except that we now have $T_u = T_w.$ Finally, we say that $X_v$ and $X_w$ are compatible and write $X_v \compatible X_w$ if 
 	\begin{itemize}
 		\item $T_v \in X_v \Leftrightarrow S_w \in X_w.$
 	\end{itemize}
 	This is because $T_v$ and $S_w$ are the same vertex of the CFG.
 	
 	In this step, our algorithm sets
 	\begin{equation*} 
 		\resizebox{\linewidth}{!}{
 		$\displaystyle \DP[u, X] = \min_{\makecell[c]{X \compatible X_v \\ X \compatible X_w \\ X_v \compatible X_w}} \DP[v, X_v] + \DP[w, X_w] - [T_v \in X_v] \cdot l(T_v) - [B_v \in X_v] \cdot l(B_v) - [C_v \in X_v] \cdot l(C_v).$}
 	\end{equation*}
 	This is because every edge in $G_u$ appears in either $G_v$ or $G_w$ but not both. Thus, the cost of the edges would simply be the sum of their costs in the two subgraphs. However, when it comes to vertices, $T_v$ and $S_w$ are merged, as are $B_v$ and $B_w,$ and $C_v$ and $C_w.$ Hence, we have to make sure we do not double count the cost of liveness for these vertices. Since this cost is counted in both $\DP$ values at the children, we should subtract it.
 	
 	\item \emph{Parallel Nodes:} We can handle parallel nodes in the same manner as series nodes, i.e.~finding compatible masks at both children and ensuring that there is no double-counting of the costs of vertices. To be more precise, let $G_u = \oparal{G_v}{G_w}.$ The compatibility conditions we have to check are as follows:
 \begin{equation*}
 	\resizebox{\linewidth}{!}{
 	$\begin{matrix}
 		X \compatible X_v \Leftrightarrow \\ \left( S_u \in X \Leftrightarrow S_v \in X_v ~\land~ T_u \in X \Leftrightarrow T_v \in X_v ~\land~ B_u \in X \Leftrightarrow B_v \in X_v ~\land~ C_u \in X \Leftrightarrow C_v \in X_v \right);\\
 		X \compatible X_w \Leftrightarrow \\ \left( S_u \in X \Leftrightarrow S_w \in X_w ~\land~ T_u \in X \Leftrightarrow T_w \in X_w ~\land~ B_u \in X \Leftrightarrow B_w \in X_w ~\land~ C_u \in X \Leftrightarrow C_w \in X_w \right);\\
 		X_v \compatible X_w \Leftrightarrow \\ \left( S_v \in X_v \Leftrightarrow S_w \in X_w ~\land~ T_v \in X_v \Leftrightarrow T_w \in X_w ~\land~ B_v \in X_v \Leftrightarrow B_w \in X_w ~\land~ C_v \in X_v \Leftrightarrow C_w \in X_w \right).
 	\end{matrix}$}
\end{equation*}
 	With the same argument as in the previous case, our algorithm sets
 \begin{equation*}\resizebox{\linewidth}{!}{$\displaystyle \DP[u, X] = \min_{\makecell[c]{X \compatible X_v \\ X \compatible X_w \\ X_v \compatible X_w}} \DP[v, X_v] + \DP[w, X_w] - [S_v \in X_v] \cdot l(S_v) -  [T_v \in X_v] \cdot l(T_v) - [B_v \in X_v] \cdot l(B_v) - [C_v \in X_v] \cdot l(C_v).$}
 	\end{equation*}
 	\item \emph{Loop Nodes:} Finally, we should handle the case where $G_u = \oloop{G_v}.$ This case is quite simple. By construction, in comparison to $G_v,$ the graph $G_u$ has four new vertices $$V_{\text{new}}=\{S_u, T_u, B_u, C_u\}$$ and five new edges $$E_{\text{new}} = \{ (S_u, S_v), (S_u, T_u), (T_v, S_u), (C_v, S_u), (B_v, T_u) \}.$$ The two graphs $G_u$ and $G_v$ do not share any special vertices, i.e.~$\varGamma_u \cap \varGamma_v = \emptyset.$ Moreover, for every edge $(x, y) \in E_{\text{new}}$ we can decide whether $(x, y)$ is in the calculation set solely based on $X$ and $X_v.$ This is because $x, y \in X \cup X_v.$ More specifically, $(x, y)$ is in the calculation set if and only if
 	$$
 	\varphi(X, X_v, x, y) := [x \not\in X \cup X_v \setminus I ~\land~ y \in U \cup X \cup X_v]
 	$$
 	
 	Thus, our algorithm sets:
 	$$
 	\DP[u, X] = \sum_{x \in V_{\text{new}} \cap X} l(x) + \min_{X_v \subseteq \varGamma_v} \DP[v, X_v] + \sum_{(x, y) \in E_{\text{new}}} 	\varphi(X, X_v, x, y) \cdot c(x, y).
 	$$
\end{itemize}

\textbf{Step 3 (Computing the Final Answer)} Let $r$ be the root of the SPL decomposition. By definition, we have $G_r = G.$ The algorithm outputs
$
	\min_{X \subseteq \varGamma_r} \DP[r, X]
$
as the minimum possible cost for the given LOSPRE input. This is because $G_r$ is the entire CFG $G$ and any solution $L$ will conform to exactly one of the different possible values of $X$ at $r.$ As is standard in dynamic programming approaches, one can reconstruct the optimal life set $L$ that leads to this minimal cost by retracing the steps of the algorithm and remembering which choices led to the optimal value at each step.

\begin{theorem} Given a LOSPRE instance consisting of a closed structured program $P,$ its control-flow graph $G$ with $n$ vertices, a use set $U,$ an invalidating set $I$ and two cost functions $c: E \rightarrow K$ and $l: V \rightarrow K,$ the algorithm above solves the LOSPRE problem in $O(n)$ and outputs 
	$$
	\min_{L} \cost(G, U, I, L, c, l) ~~~~~~ \text{ and } ~~~~~~
	\argmin_{L} \cost(G, U, I, L, c, l).
	$$
\end{theorem}
\textbf{Proof:}
	Correctness has already been argued above. Thus, we focus on the runtime analysis. The SPL decomposition has $O(n)$ vertices and can be computed in $O(n)$. At each vertex $u$ of the decomposition, we have $2^4 = 16 = O(1)$ different possible values for $X.$ The computations in the atomic node are over graphs with only four vertices and thus take $O(1)$ time. In a series and parallel node, we have at most two compatible $X_v$'s for each $X.$ This is because inclusion or exclusion of the vertices $S_v, B_v$ and $C_v$ in $X_v$ is uniquely determined by $X$ and only $T_v$ remains to be chosen. Similarly, for every fixed $X, X_v,$ there is a unique $X_w.$ Thus, computing each $\DP[u, X]$ in this step takes $O(1)$ time. In a loop node, every $X$ induces a unique $X_v$ and a unique $X_w.$ Hence, this step takes $O(1)$ time to compute each $\DP[u, x]$ value. In step 3, we try $2^4 = O(1)$ different $X_v$'s for each $X.$ Thus, the total runtime of Step 2 is $O(n).$ Finally, Step 3 takes the maximum of $2^4 = O(1)$ values.

\newpage

\chapter{Partial Constraint Satisfaction Problem}
\label{chp_background}
\paragraph{Constraint Satisfaction Problem.}The famous Constraint Satisfaction Problem (CSP)\cite{PCS} is defined as a tuple \(\langle V, D, C\rangle\), where \(V\) represents a set of variables, \(D\) is the domain set for all \(v \in V\), and \(C\) is a set of constraints. A CSP is solved by finding an assignment of values to the variables that satisfies all constraints. When applied to graphs, we treat each node as a variable and each edge as a constraint, with the stipulation that constraints exist only between adjacent nodes. For example, the graph coloring problem can be formulated as a CSP where each node is a variable, and each edge imposes a constraint that the colors of adjacent nodes must differ. In this scenario, we aim to assign a color to each node such that no two adjacent nodes share the same color. It is well-known that the graph coloring problem is NP-hard even when the domain set is limited to only colors, which implies that the CSP problem is also NP-hard.

\paragraph{Partial Constraint Satisfaction Problem.}  In the context of binary relationship PCSPs (Parameterized Constraint Satisfaction Problems) \cite{PCS}, we allow certain constraints to be violated at a specified cost, with the goal of finding a solution that minimizes this cost. To define the cost, we introduce a cost function \(c(e, b_0, b_1)\), where \(e\) is the edge, and \(b_0\) and \(b_1\) are the values assigned to the two nodes connected by the edge. If \(b_0\) and \(b_1\) do not violate the constraints, the cost is 0; otherwise, a positive cost is assigned. Our objective is to find:

\[
\min_{A} \sum_{e \in E} c(e, A(v_0), A(v_1))
\]

where \(A: V \to D\) maps each node to a domain.

If we assign an infinite cost to the constraints, the problem reduces to the CSP, confirming its NP-hardness.

\paragraph{Tree-Decomposition based solution.}
This section is mainly followed the algorithm from \cite{Koster2002SolvingCS}.

The algorithm is founded on the following concept: Let \( S_V \) be a separating vertex set of \( G \) such that \( G[V \setminus S] = G[V_1] \cup G[V_2] \). In this case, the optimal assignment in \( V_1 \) (or \( V_2 \)) depends solely on the assignment in \( S \). Thus, given an assignment for \( S \), the problem decomposes into two independent PCSPs on \( G[V_1] \) and \( G[V_2] \), which can be solved separately. This concept can be expressed as a dynamic programming algorithm utilizing a tree decomposition \( (T, \mathcal{B}) \) of the graph. For every internal node \( i \in I \), \( X_i \) represents a separating vertex set, indicating that, given an assignment for \( X_i \), the PCSP decomposes into smaller PCSPs for each branch in the tree, and hence develop a parameterized algorithm with treewidth as the parameter.

Given that the treewidth of a goto-free structured program is at most 7, we can decompose the PCSP problem into \( O(G) \) smaller PCSPs, each containing at most 8 nodes, allowing us to solve them in linear time. Many compiler optimization tasks, including Register Allocation \cite{RegisterAllocation}, LOSPRE \cite{LOSPRE}, and the placement of Bank Selection Instructions \cite{bank_selection}, rely on this approach. However, as previously mentioned, tree decomposition does not leverage the sparsity of the control flow graph (CFG). Furthermore, tree decomposition treats the CFG as an undirected graph, which results in the loss of directional information and limits the ability to perform specific optimizations. Our SPL-decomposition addresses these limitations.

\paragraph{General Solution with SPL-decomposition.}
Our algorithm proceeds with a bottom-up dynamic programming on the SPL decomposition. Note that each node $u$ of the SPL decomposition corresponds to an SPL subgraph $G_u = (V_u, E_u, S_u, T_u, B_u, C_u)$ of $G$ which is either an atomic SPL graph (when $u$ is a leaf) or obtained by applying one of the SPL operations to the graphs corresponding to the children of $u.$. Let $\varGamma_u = \{S_u, T_u, B_u, C_u\}$ be the set of special vertices of $G_u$. Let $X$ be the assignment for the $\varGamma_u$. We define a dynamic programming variable $\DP[u, X].$ Our goal is to compute this dynamic programming value such that

$$
\DP[u, X] = \min_{A|X}\sum_{e \in E_u} c(e, A(v_0), A(v_1))
$$

where the minimum is taken over all assignments $A$ to the vertices of $G_u$ that agree with $X$ on the special vertices. In other words, $A(v) = X(v)$ for all $v \in \varGamma_u$.

\begin{enumerate}
    \item \emph{Atomic Nodes}: If $G_u$ is an atomic SPL graph, then the only vertices in $G_u$ are the four special vertices. Therefore, we must have $A=X$ Our algorithm computes each $\DP[u, X]$ as:
	 $$\DP[u, X] = \sum_{e \in E_u} c(e, X(v_0), X(v_1))$$

    \item \emph{Series Nodes}: Suppose $G_u = \oseries{G_v}{G_w}$ where $v$ and $w$ are the children of $u$ in the SPL decomposition. 
    Let $X$ be the assignment of special nodes in $G_u$, $X_v$ be the assignment of special nodes in $G_v$, and $X_w$ be the assignment of special nodes in $G_w$. 
     We say that $X_v$ and $X_w$ are compatible and write $X \compatible X_v$ if the following conditions satisfied:
    \begin{itemize}
    \item 	$X(S_u)=X_v(S_v);$
    \item	$X(B_u)=X_v(B_v);$
    \item	$X(C_u)=X_v(C_v).$
    \end{itemize}
    Intuitively, compatibility means that the assignments $X_v$ and $X$ return the same value when given the same vertex.
    
    Now consider $X_w$ and $X$. We say that $X_w$ and $X$ are compatible and write $X \compatible X_w$ if the following conditions are satisfied:
    \begin{itemize}
        \item 	$X(T_u)=X_w(T_w);$
        \item	$X(B_u)=X_w(B_w);$
        \item	$X(C_u)=X_w(C_w).$
        \end{itemize}
    The intuition is the same as the previous case, except that we now have $T_u = T_w.$ Finally, we say that $X_v$ and $X_w$ are compatible and write $X_v \compatible X_w$ if 
    \begin{itemize}
        \item $X_v(T_v)=X_w(S_w).$
    \end{itemize}
    This is because $T_v$ and $S_w$ are the same vertex of the CFG.
    
    In this step, our algorithm sets
    $$
	\DP[u, X] = \min_{\makecell[c]{X \compatible X_v \\ X \compatible X_w \\ X_v \compatible X_w}} \DP[v, X_v] + \DP[w, X_w].
    $$
    This is because every edge in $G_u$ appears in either $G_v$ or $G_w$ but not both. Thus, the cost of the edges would simply be the sum of their costs in the two subgraphs.
    
    \item \emph{Parallel Nodes:} We can handle parallel nodes in the same manner as series nodes, i.e.~finding compatible masks at both children. 
    To be more precise, let $G_u = \oparal{G_v}{G_w}.$ The compatibility conditions we have to check are as follows:
    $$
    \begin{matrix}
        X \compatible X_v \Leftrightarrow \left( X=X_v \right);\\
        X \compatible X_w \Leftrightarrow \left( X=X_w \right);\\
        X_v \compatible X_w \Leftrightarrow \left( X_v=X_w \right).
    \end{matrix}
    $$
    As the special nodes set in the three SPL nodes should be the same.

    With the same argument as in the previous case, our algorithm sets
    $$
    \DP[u, X] = \min_{\makecell[c]{X \compatible X_v \\ X \compatible X_w \\ X_v \compatible X_w}} \DP[v, X_v] + \DP[w, X_w].
    $$
    \item \emph{Loop Nodes:} Finally, we should handle the case where $G_u = \oloop{G_v}.$ 
    By construction, in comparison to $G_v,$ the graph $G_u$ has four new vertices 
    $$V_{\text{new}}=\{S_u, T_u, B_u, C_u\}$$ and three new edges 
    $$E_{\text{new}} = \{ (S_u, S_v), (S_u, T_u), (T_v, S_u)) \}.$$ 
    As  $C_v$ and $S_u$ should keep the same value, as well as $B_v$ and $T_u$, the compatibility to be checked are as follows:
    $$
        X \compatible X_v \Leftrightarrow \left( X(S_u)=X_v(C_v) \right) \text{AND} \left( X(T_u)=X_v(B_v)  \right);\\
    $$

    Also, knowing $X$ and $X_v$ is enough to calculate the cost of the new edges. Let's call the cost of the new edges based on $X$ and $X_v$ $c(e)$.
    
    Thus, our algorithm sets:
    $$
    \DP[u, X] = \min_{X \compatible X_v} \DP[v, X_v] + \sum_{e \in E_{\text{new}}}(c(e)).
    $$
\end{enumerate}

After finishing the dynamic programming, the minimum cost is given by $min_X \DP[root, X]$ where $root$ is the root of the SPL decomposition and $X$ is the assignment of the special nodes of the root. The exact assignment can be easily tracked and updated during the dynamic programming.

 Let's analyze the time complexity of each dynamic programming step.
\begin{itemize}
    \item \textbf{Atomic node:} As we only have four nodes inside the atomic nodes, and each of them has $|D|$ possible values,
    the time complexity is \(O(|D|^4)\). To optimize the time complexity, we can only consider the connected nodes, and for the unconnected nodes, we can temporarily ignore them as there is no constraint on them now. As there are at most three connected nodes in atomic nodes, the time complexity is \(O(|D|^3)\).
    \item \textbf{Series node:} For series nodes, there are totally eight special nodes that need to be considered, but as we also need to consider the compatibility, there are three pairs among them that need to have the same value. Thus, the time complexity is
    \(O(|D|^5)\). As we only need to consider the value for five different variables. 
    \item \textbf{Parallel node:} Similar to the series node, but there are four pairs of nodes that need to be with the same value. Thus, the time complexity is $O(|D|^4)$
    \item \textbf{Loop node:} Similar to the series and parallel nodes, but there are two pairs of nodes that need to be with the same value. Thus, the time complexity is $O(|D|^6)$. However, as we know that the $C_u$ and $B_u$ are not connected after the loop operation, we can temporarily ignore them, and hence the time complexity is $O(|D|^4)$.
\end{itemize}
In all, all steps in the dynamic programming can be done in $O(|D|^5)$, and the size of the SPL-decomposition is polynomial to the size of CFG, hence the overall time complexity is $O(|G|\cdot|D|^5)$.

\paragraph{Register Allocation.}
In this case, the "value" assigned to each node is the allocation of an alive variable. Suppose there are at most $V$ alive variables at one node and there are $r$ registers, then we can create a completed inference graph with $V$ nodes and try to color them with $r$ colors. Then the domain size is $$\binom{V}{r} \cdot r! + \binom{V}{r-1} \cdot (r-1)! + \dots + \binom{V}{0} \cdot 0! \in O(r \cdot V^r).$$ Hence with our algorithm, the time complexity is $O(|G| \cdot r^5 \cdot V^{5 \cdot r}).$ 

There is a specific version of the register allocation problem named Spill-free Register Allocation. In this case, $c(e, a_1, a_2) = 0$ if the two assignments are valid and allocates all variables to registers meaning that it does not map anything to $\perp$ and $c(e, a_1, a_2) = +\infty$ otherwise, and simply asking whether an assignment with zero total cost is attainable. In this case if we have $V > r$, then we can directly answer "no", hence we can have a domain with size $O( r^{r+1})$, which means our algorithm is a XP algorithm with parameter $r$ and with time complexity  $O(|G| \cdot r^{5 \cdot r + 5})$ for this case. 

\paragraph{LOSPRE.}
To consider this problem as PCSP, we only need to have a little modification, as the edge cost is the same as the definition in PCSP, and the node cost can be easily added to the total cost. In this case, the "value" assigned to each node is if it is belong to Use set, Life set, and Invalidating set, as for each set, there are 2 different case, hence the domain size is $2^3=8$, and hence, applying to our PCSP algorthm, this can be down in $O(|G|\cdot 8^5)=O(|G|)$, hence in this case, thanks to the constant domain size, we can develop a linear algorithm without taken any parameter.

\newpage

\chapter{Placement of Bank Selection Instruction}
\label{chp_background}
Partitioned memory architectures\cite{10.1145/1176760.1176786} are prevalent in 8-bit and 16-bit microcontrollers. In these systems, a portion of the logical address space serves as a window into a larger physical address space. The segments of the physical address space that can be mapped into this window are referred to as memory banks. A mechanism exists to determine which part of the physical address space is visible within the window, typically achieved through bank selection instructions.

The assignment of variables to specific memory banks is generally performed by the programmer (for instance, through named address spaces in Embedded C) or by the compiler (using techniques such as bin-packing heuristics to minimize RAM usage). Some approaches integrate the placement of variables in memory banks with the insertion of bank selection instructions. However, in embedded systems, there is often more available space for code than for data. As a result, variables stored in banked memory tend to be larger, making it advantageous to prioritize the efficient packing of variables into the banks before addressing other factors such as code size and execution speed. Consequently, the placement of variables in memory typically occurs at an earlier stage of the compilation process than the insertion of bank-switching instructions\cite{bank_selection}.

Let $D$ be the memory bank domain, including a special symbol \(\perp \in D\) that indicates that the currently selected bank is unknown. A program can be modeled as a control-flow graph \( G = (V, E) \), where \( E \subseteq V^2 \), and each node \( v \in V \) can be assigned a memory bank from $D$. Some of the nodes are percolored as the specific bank must be active at that node.
\begin{figure*}
	\centering
    \noindent\makebox[\textwidth]{
	\subfloat[Program]{
		\includegraphics[clip,width=.33\textwidth]{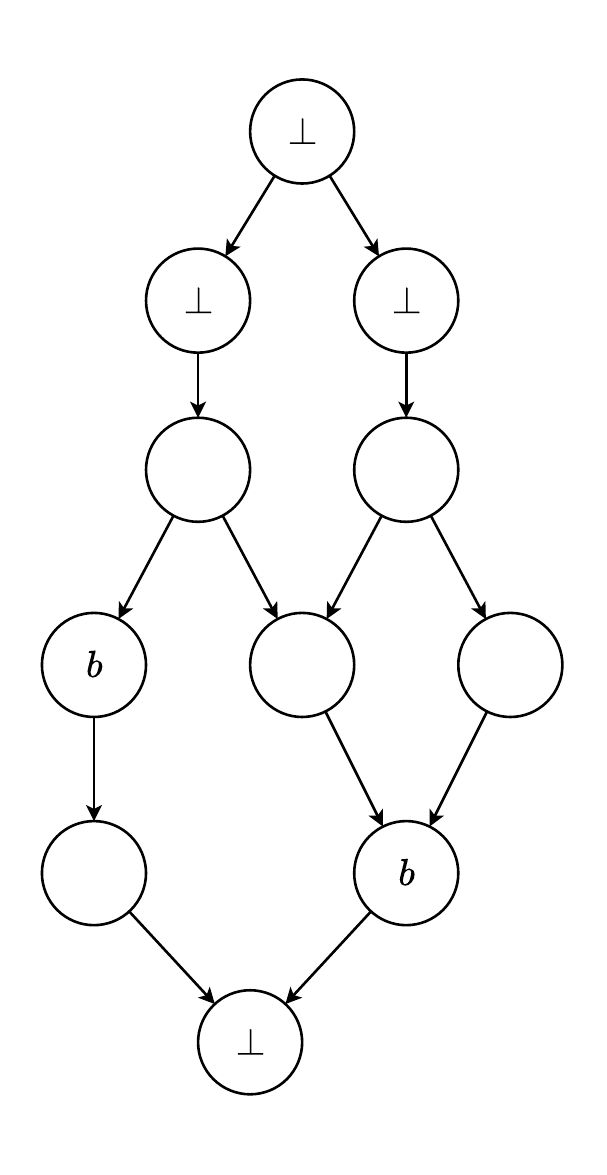}
	}%
	\subfloat[Ad-hoc Aproach]{
		\includegraphics[clip,width=.33\textwidth]{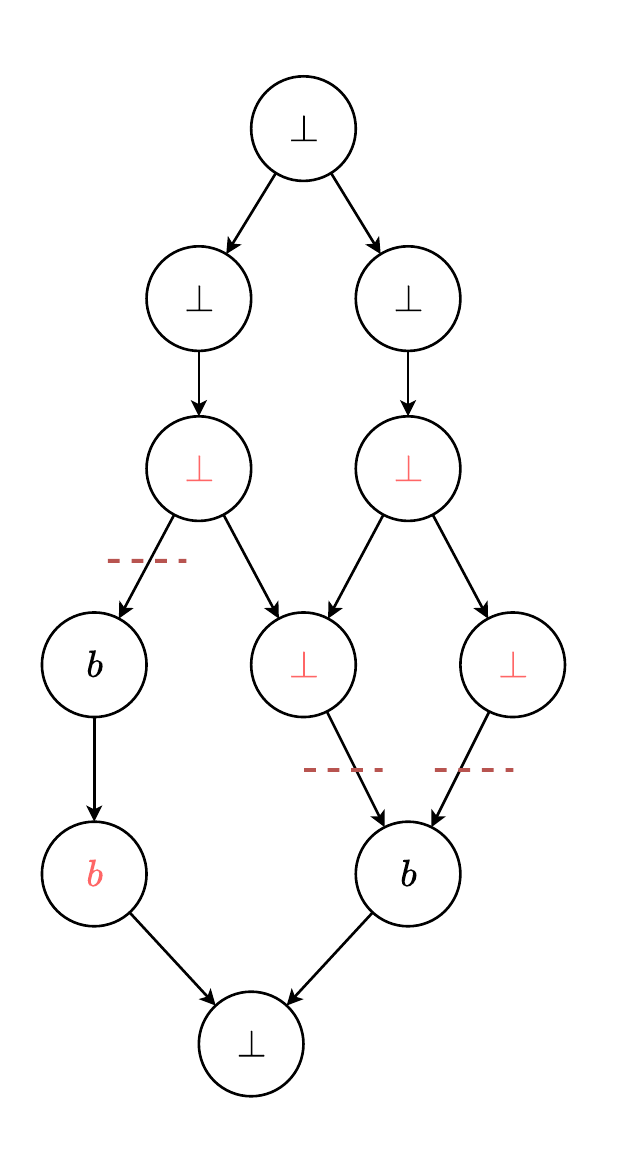}
	}%
	\subfloat[Optimal]{
		\includegraphics[clip,width=.33\textwidth]{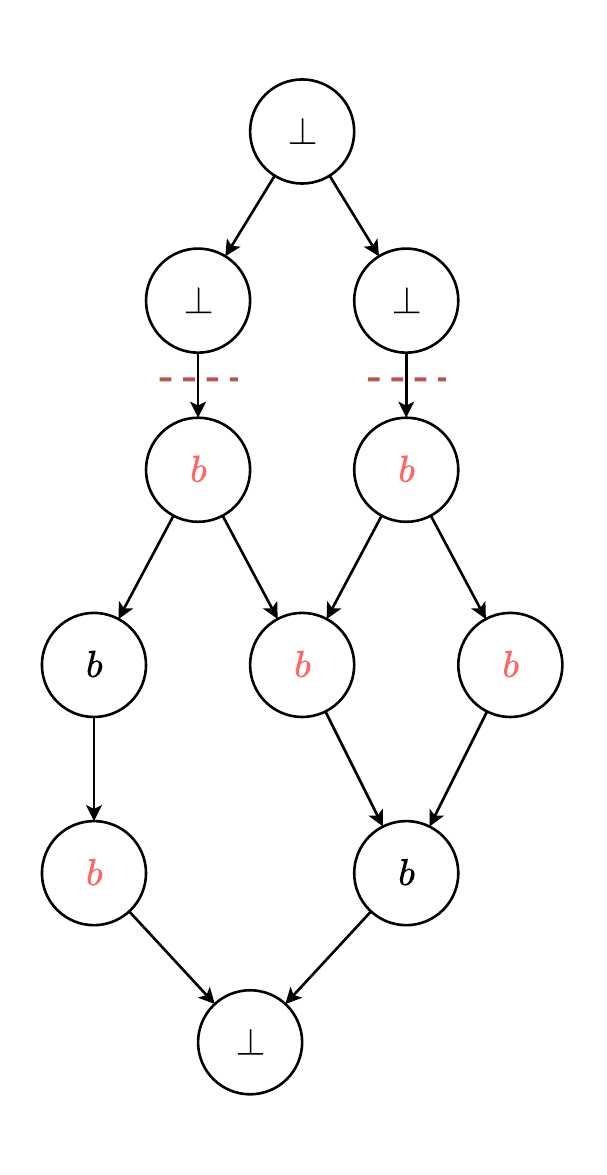}
	}
    }
    \caption{Example Instance}
	\label{fig:bank-selection}
\end{figure*}

Fig~\ref{fig:bank-selection} shows an example of a program with bank selection instructions. The program has three percolored node and the bank $b$ should be active at these nodes. For the other uncolored nodes, that means the instructions there do not need to access the bank memory; hence, which bank is active does not matter. For the simple ad-hoc approach, we can add the bank selection instruction just before we need the bank to be active, like Fig~\ref{fig:bank-selection} (b), which needs to insert three instructions. In this case, suppose we only want to minimize the number of bank selection instructions, we can insert the bank selection instruction at the beginning of the program like Fig~\ref{fig:bank-selection} (c), which only needs to insert two instructions, and which is the optimal solution.

A cost function for a control flow graph \( G = (V, E) \) is a function \( c: E \times D \times D \to \mathbb{R} \) that assigns a cost to each edge \( e \in E \) based on the memory banks assigned to its endpoints. In other words, for $c(e,b_0,b_1)$, it means the cost if $b_0$ is active before the edge e and $b_1$ is active after the edge.

Cost function can be designed based on different optimization criteria. A typical cost function for optimizing code size can be defined as follows: 

\begin{itemize}
    \item \( c(e, b, b) = c(e, b, \perp) = 0 \) since no instructions need to be inserted. 
    \item \( c(e, b_0, b_1) = c_1 > 0 \) for \( b_0 \neq b_1 \neq \perp \) when \( e \) is an edge from a taken conditional branch, as splitting such an edge generates an additional unconditional jump instruction. 
    \item For all other cases, \( c(e, b_0, b_1) = c_0 > 0 \) for \( b_0 \neq b_1 \neq \perp \), with the condition that \( c_0 < c_1 \).
\end{itemize}

The goal is to find an assignment of memory banks to the nodes of the control flow graph that minimizes the total cost of the edges. In other words, we want to find an assignment A such that
$$
\min_{A} \sum_{e \in E} c(e, A(v_0), A(v_1))
$$
where \( A: V \to D \) maps each node to a domain.

It is important to note that this problem is NP-hard, even when the cost function is simplified to \( c(e, b_0, b_1) = 0 \) for \( b_0 = b_1 \) or \( b_1 = \perp \), and \( c(e, b_0, b_1) = 1 \) in all other cases.

 It is easy to find that this problem is a typical PCSP graph problem and can easily apply the general solution we mentioned in the previous chapter and solved in $O(|G|\cdot|D|^5)$ time complexity, where here $|D|$ is the size of possible banks.

\newpage

\chapter{Experiments}
\label{chp_background}
In this section, we provide experimental results comparing my algorithm for spill-free register allocation, LOSPRE, and optimization on placement of bank selection instructions with previous approaches based on treewidth\cite{DBLP:conf/cc/Krause13, LOSPRE, bank_selection}. As for all three tasks, both approaches can get an optimal solution, We only compare their runtime. Additional related experiment results can be found at the end of this section.

\paragraph{Implementation.} We implemented our approach in \texttt{C++} and integrated it with the Small Device C Compiler (SDCC)~\cite{sdcc,sdcc2}. SDCC already includes a heavily optimized variant of the algorithms from~\cite{DBLP:journals/iandc/Thorup98,DBLP:journals/dam/KrauseLS20} for finding tree decompositions and the treewidth-based algorithm for the three tasks. Despite our approach being perfectly parallelizable, we did not use parallelization in our experiment in order to provide a fair comparison with the available implementations of previous methods, which are not parallel. 

\paragraph{Machine.}  The results were obtained on a virtual machine with Oracle Linux (ARM 64-bit), equipped with 1 core CPU of Apple M2 and 4GB of RAM.

\begin{figure}
\centering
	\includegraphics[width=0.6\linewidth]{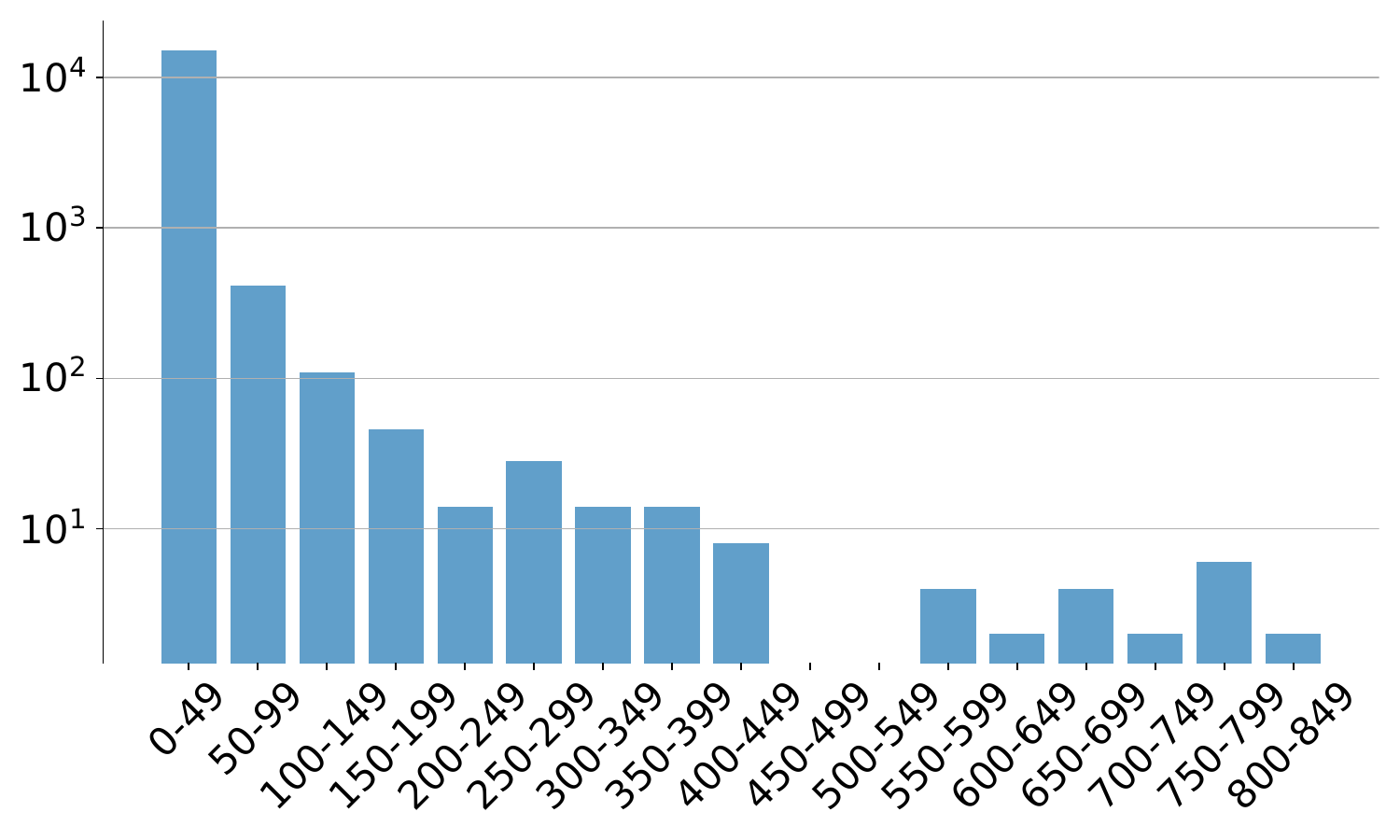}
	\caption{Histogram of the number of CFG vertices (lines of code).}
	\label{fig:cfgSize}
\end{figure}

\paragraph{Benchmark.}We followed the setup described in~\cite{conradof2023bounded}. We used the SDCC regression test suite for HC08 as our benchmark set. These benchmarks consist of embedded programs that are designed to run on systems with limited resources, making compiler optimization a critical performance bottleneck for them. The functions within these benchmarks have control flow graphs (CFGs) ranging from 1 to 800 vertices (lines of code), with an average size of 15.7 vertices. Figure~\ref{fig:cfgSize} presents a histogram of function sizes. The $x$ axis is the CFG size, and the $y$ axis is the number of instances. The $y$ axis is on a logarithmic scaleWe established a time limit of 10 minutes for the tests. While some instances may not require specific optimization and were thus excluded from the related tests, We collected over 15,000 valid instances for each of the three cases.

\begin{figure}
\centering
	\includegraphics[width=0.44\linewidth]{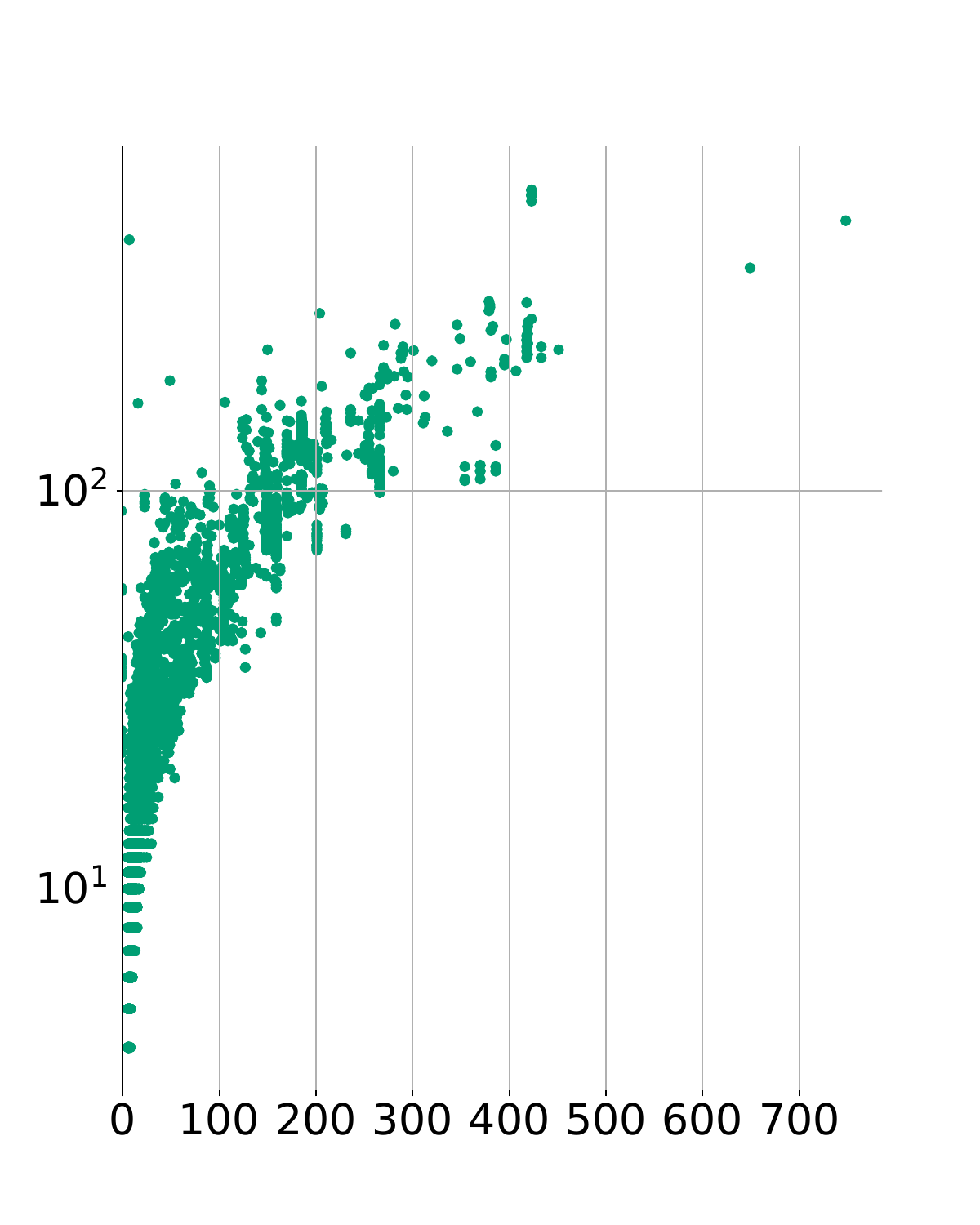}
	\caption{The runtime needed for computing grammatical decompositions of CFGs by parsing the programs. }
	\label{fig:spl-decom-time}
\end{figure}

\begin{figure}
\centering
	\includegraphics[width=0.6\linewidth]{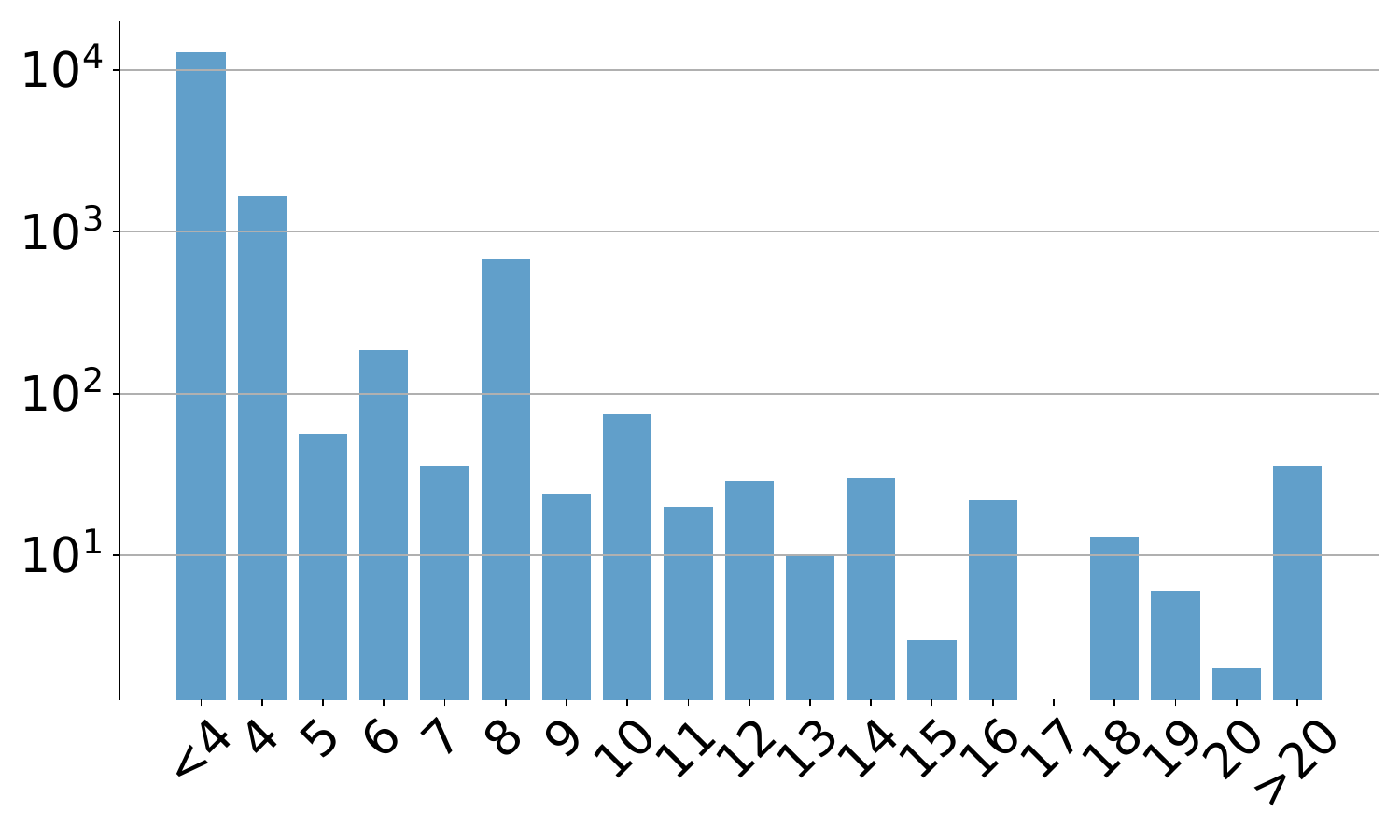}
	
	\caption{Histogram of the minimum number of registers required for spill-free allocation. }
	\label{fig:register-frequency}
\end{figure}

\paragraph{Runtimes for Computing the Grammatical Decomposition.} All three compiler optimization tasks rely on grammatical decompositions of control flow graphs (CFGs) as SPL graphs. As mentioned in Chapter 3, such decompositions can be computed in linear time with a single parse of the program. Figure~\ref{fig:spl-decom-time} illustrates the runtime required for computing grammatical decompositions for each of our benchmarks. Each dot corresponds to one instance. The $x$ axis is the size of the CFG, and the $y$ axis is the runtime in microseconds. The $y$ axis is on a logarithmic scale. The average runtime was 13.8 microseconds, with a maximum of 570 microseconds. Therefore, grammatical decompositions can be computed extremely efficiently, and the time spent obtaining them does not significantly contribute to the total compile time.

\begin{figure}
\centering
	\includegraphics[width=.5\linewidth]{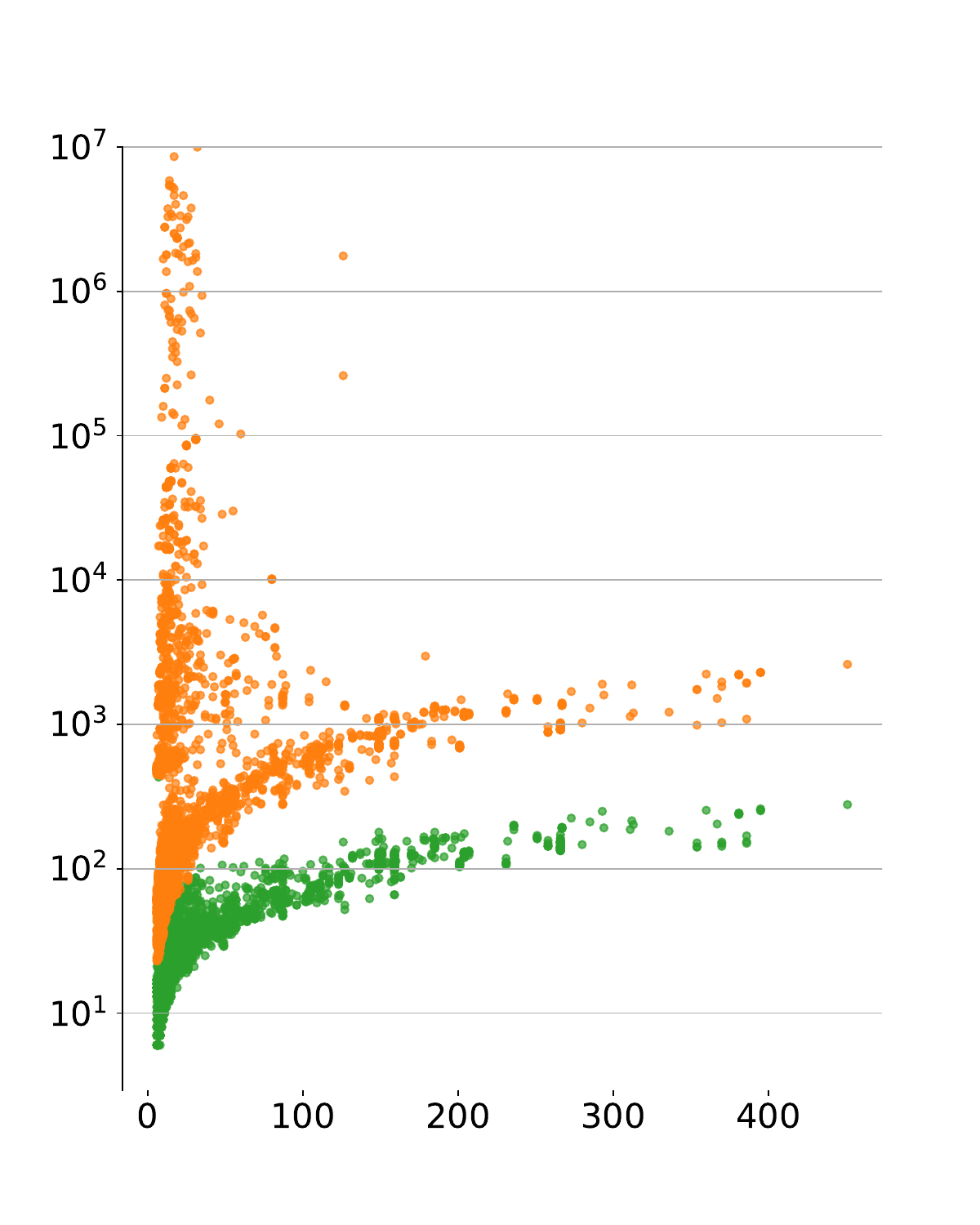}
	\caption{Runtime comparison of the treewidth-based algorithm (orange) and our approach (green) for Register Allocation}
	\label{fig:runtime-ra}
\end{figure}

\paragraph{Register Allocation. }
Specifically, given an input program, our objective is to determine the smallest number \( r \) of registers required for spill-free allocation. As previously mentioned, spill-free register allocation is a specific case of register allocation characterized by minimum cost, where the cost is zero if there is no spilling and infinite otherwise. We selected this problem for our experimental evaluation for two reasons: (i) most previous works in the literature focus on this variant, and (ii) there is generally no standard method for selecting the cost function \( c \); each compiler defines this function differently based on its own context and use cases, often relying on dynamic analysis and profiling. In contrast, spill-free allocation is well-defined and consistent across all compilers.

 Our approach successfully handled all input instances within the prescribed time and memory limits, either finding the optimal number of registers needed for spill-free allocation or reporting that more than 20 registers are required. Figure~\ref{fig:register-frequency} shows a histogram of the number of required registers. The $x$ axis is the number of registers, and the $y$ axis is the number of instances requiring that many registers. The $y$ axis is on a logarithmic scale. In contrast, the treewidth-based approach of~\cite{DBLP:conf/soda/BodlaenderGT98} failed in 554 instances, including all instances requiring more than 8 registers.

Figure~\ref{fig:runtime-ra} shows a comparison of the runtimes of our algorithm vs the treewidth-based approach of~\cite{DBLP:conf/soda/BodlaenderGT98}. The $x$ axis is the number of vertices in the CFG, and the $y$ axis is the time in microseconds; the $y$ axis is on a logarithmic scale. When we set $r \leq 20.$, the average runtimes were 3.87 microseconds for our approach and  1,191,284  microseconds for~\cite{DBLP:conf/soda/BodlaenderGT98}. These averages are excluding the instances over which the previous methods failed. The runtimes were dominated by 704 instances for the treewidth-based approach, presumably due to high treewidth. Excluding these outlier instances, the average runtime was 372.34 microseconds for~\cite{DBLP:conf/soda/BodlaenderGT98}

\begin{figure}
\centering
	\includegraphics[width=.5\linewidth]{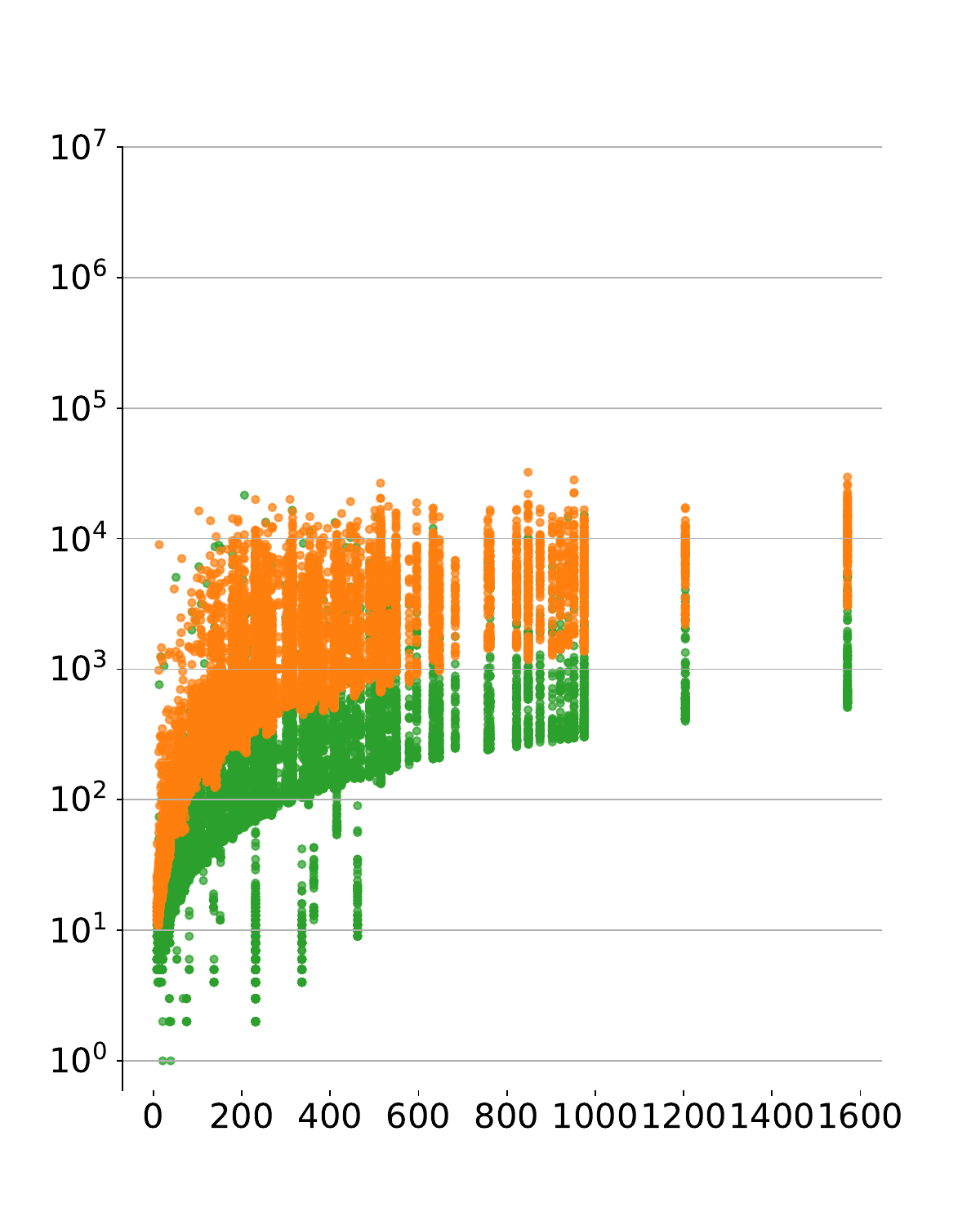}
	\caption{Runtime comparison of the treewidth-based algorithm (orange) and our approach (green) for LOSPRE}
	\label{fig:runtime-LE}
\end{figure}

\paragraph{LOSPRE.}
The goal is to minimize the total number of computations in the resulting 3-address code. Thus, we use $K = \mathbb Z^2$ with lexicographic ordering. The cost assigned to each edge $(x, y)$ is $c(x, y) = (1, 0).$ We also enforce lifetime-optimality by assigning the cost $l(x) = (0, 1)$ to every vertex $x.$

Figures~\ref{fig:runtime-LE} provide runtime comparisons between \cite{LOSPRE} and our approach. The $x$ axis is the number of vertices in the CFG, and the $y$ axis is the time in microseconds; the $y$ axis is on a logarithmic scale. On average, our algorithm takes 222.38 microseconds, while the treewidth-based approach of \cite{LOSPRE} has an average runtime of 1349.14 microseconds. The maximum runtime was 21,524 microseconds for our algorithm compared to 32,284 microseconds for \cite{LOSPRE}. Our algorithm significantly outperforms~\cite{LOSPRE} in the vast majority of benchmarks. We identified only 19 instances where our runtime exceeded 10,000 microseconds, whereas \cite{LOSPRE} takes more than 10,000 microseconds in 277 instances.

\begin{figure}
\centering
	\includegraphics[width=.5\linewidth]{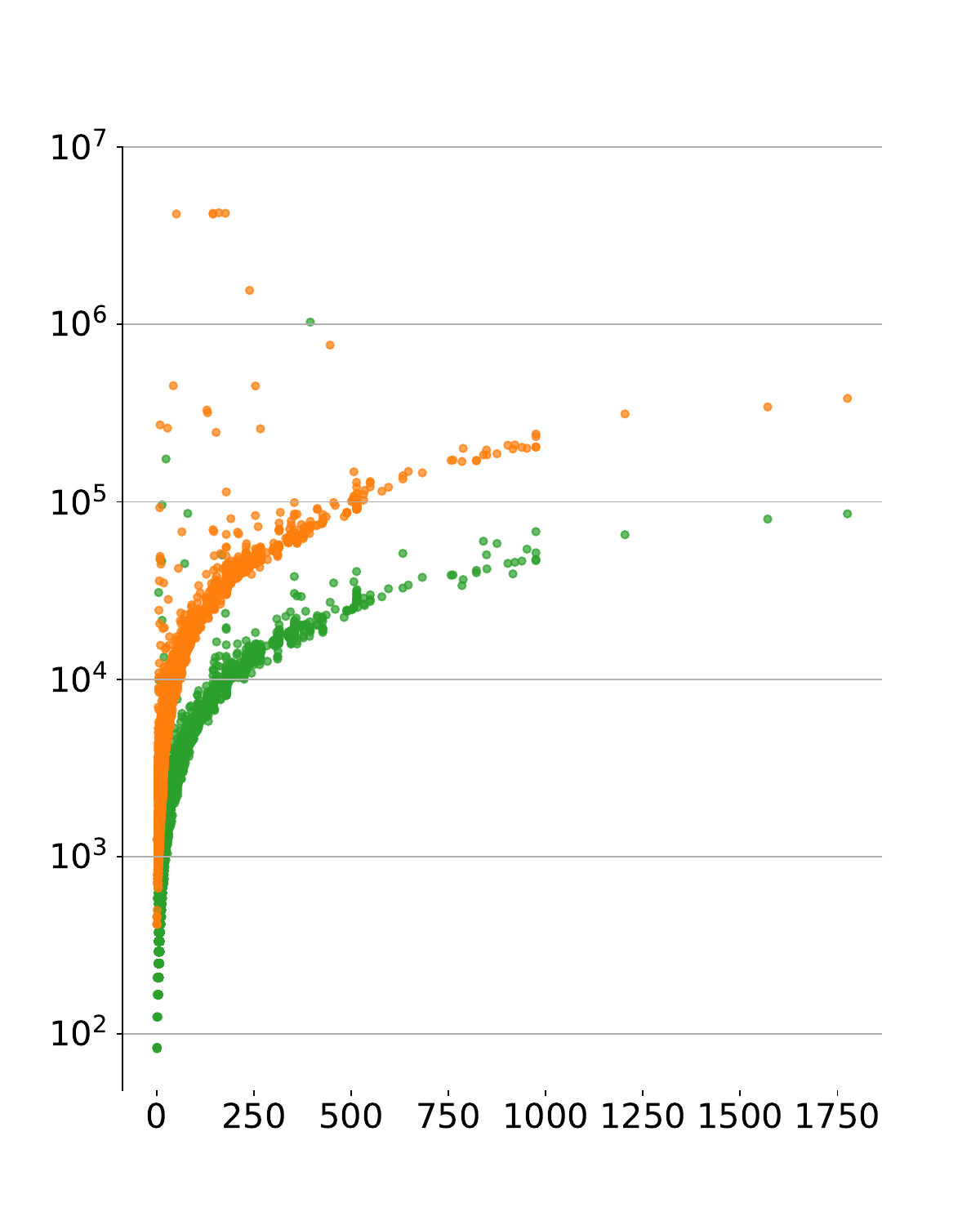}
	\caption{Runtime comparison of the treewidth-based algorithm (orange) and our approach (green) for Placement of Bank Selection Instruction}
	\label{fig:runtime-BS}
\end{figure}

\paragraph{Placement of Bank Selection Instructions.} In this experiment, we focus on optimization on code size, hence with the following cost function.
\begin{itemize}
    \item \( c(e, b, b) = c(e, b, \perp) = 0 \) .
    \item \( c(e, b_0, b_1) = 6 \)  when \( e \) is an edge from a taken conditional branch.
    \item For all other cases, \( c(e, b_0, b_1) = 3 \).
\end{itemize}

Figure~\ref{fig:runtime-BS} shows runtime of our approach and the approach of \cite{bank_selection}. The $x$ axis is the number of vertices in the CFG, and the $y$ axis is the time in nanoseconds; the $y$ axis is on a logarithmic scale. According to the data our algorithm took an average of $ 1998.3$ nanoseconds, while \cite{bank_selection} took an average of $8558.8$ nanoseconds.

\paragraph{Discussion.} In all three cases, our algorithm outperforms the previous state-of-the-art algorithm. For register allocation, our approach is the first exact algorithm for spill-free register allocation that scales to realistic architectures with up to 20 registers, such as those in the x86 family. Given the efficiency of my method, which achieves an average runtime of merely 4 microseconds per instance, we believe there is no longer a justification for using approximations or heuristics in spill-free register allocation. Despite its NP-hardness and theoretical hardness of approximation, our method efficiently solves this problem for all practical instances. For LOSPRE and the optimization of bank selection instruction allocation, our approach is at least four times faster than the treewidth-based algorithm. Considering that the treewidth-based algorithm is already efficient, this represents a significant improvement.

We intuitively believe that the source of these practical enhancements lies in the fact that our algorithm operates with smaller cuts, with a maximum size of 4, in the control flow graph (CFG) when solving the PCSP. In contrast, the treewidth-based approach utilizes cuts with up to 8 vertices. Additionally, while tree decomposition treats the CFG as an undirected graph, our method retains direction information, allowing for specific optimizations during implementation.

\begin{figure}
	\centering
		\includegraphics[clip,width=.5\textwidth]{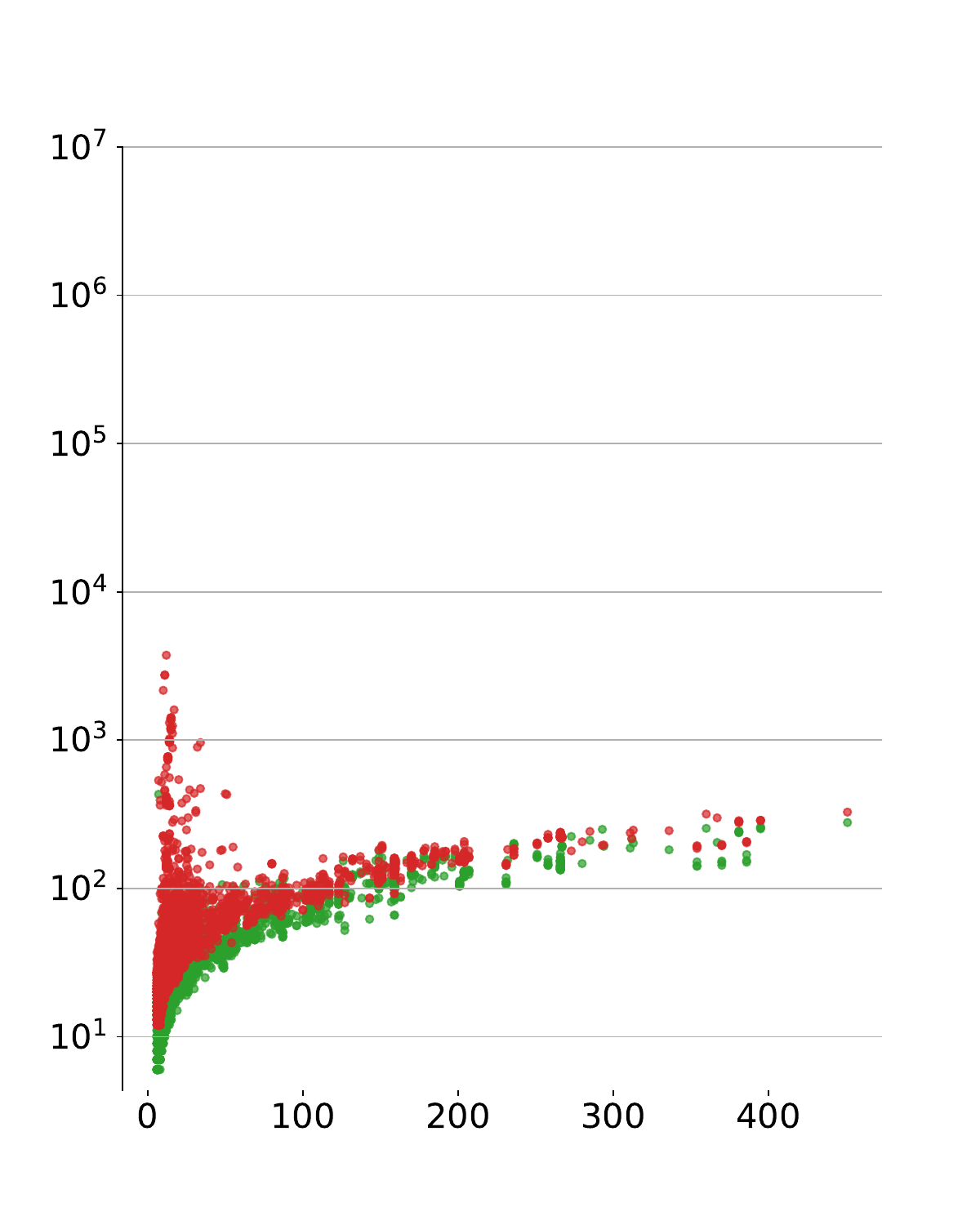}
	
	\caption{Runtime comparison of our approach with \cite{conradof2023bounded} for register allocation. }
	\label{fig:runtime-f1}
\end{figure}

\paragraph{Additional Experiments for Register Allocation.}

We further compared our approach with path decomposition based algorithm \cite{conradof2023bounded} and traditional graph coloring based algorithm \cite{DBLP:conf/pldi/Chaitin82}.

Path decomposition based \cite{conradof2023bounded} cannot handle cases with more than 20 registers either. For the valid case, the average time of \cite{conradof2023bounded} is 21,544 microseconds, which is faster than the treewidth-based algorithm we mentioned in Chapter 6, but still slower than our approach, which has an average of 3.87 microseconds. The runtime is shown in Figure~\ref{fig:runtime-f1}. The $x$ axis is the number of vertices in the CFG, and the $y$ axis is the time in microseconds. The $y$ axis is on a logarithmic scale.

\begin{figure}
	\centering
		\includegraphics[clip,width=.5\textwidth]{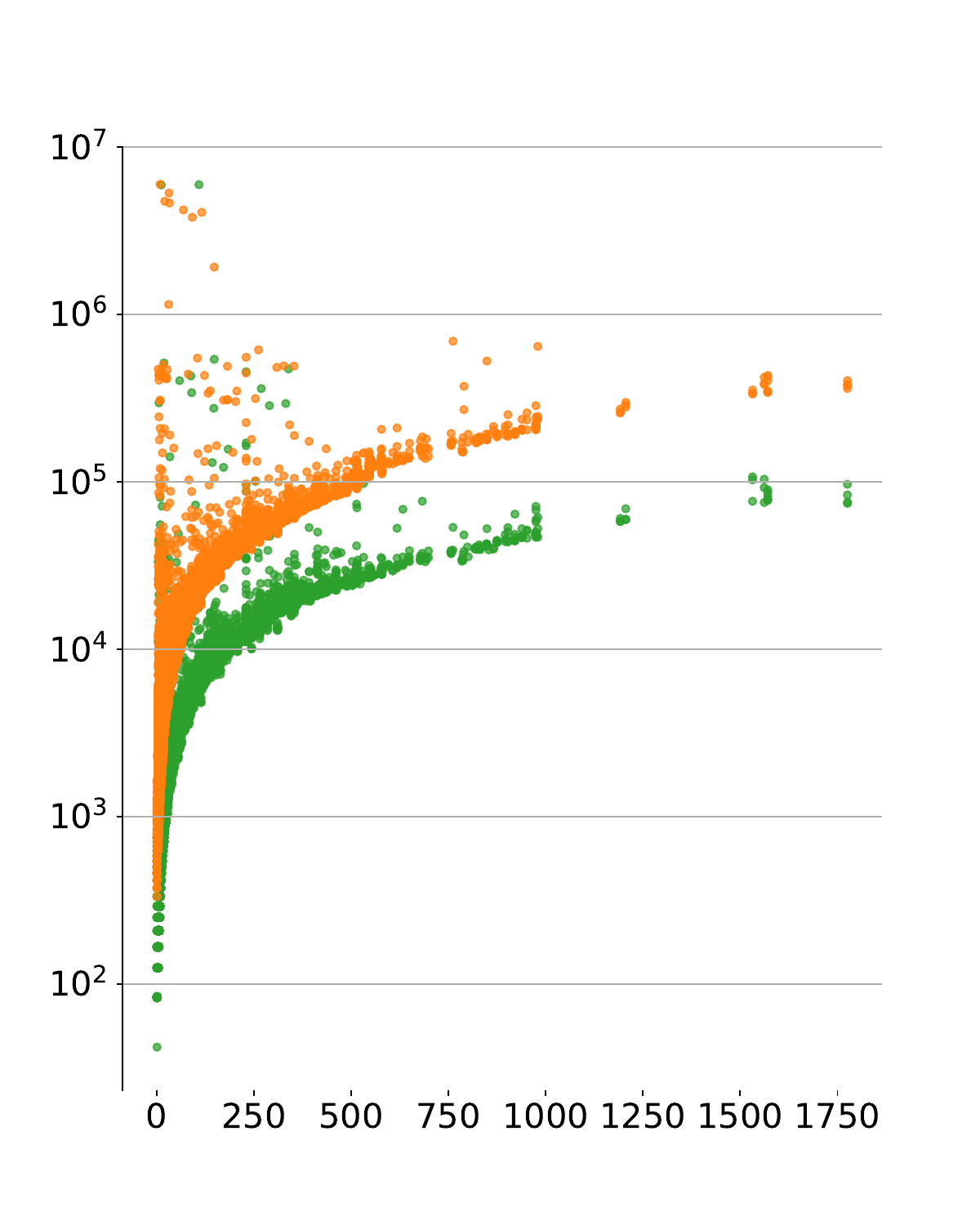}
	
	\caption{Runtime comparison of our approach with the treewidth-based algorithm \cite{bank_selection} over MC51.}
	\label{fig:runtime-f2}
\end{figure}

We observe that Chaitin's graph coloring method, which is the only classical non-parameterized approach that uses the optimal number of registers, is highly unscalable. Given a time limit of 1 minute, it could handle only 6,042 benchmarks in our suite with an average runtime of 153,602 microseconds. Notably, this did not include any benchmark that required more than $8$ registers. Graph coloring timed out on all such benchmarks. The runtime distribution is reported in Figure~\ref{fig:runtime-f2}. The $x$ axis is the number of vertices in the CFG, and the $y$ axis is the time in microseconds. The $y$ axis is on a logarithmic scale.  In comparison, our approach handles all benchmarks and has an average runtime of 3.87 microseconds.

\paragraph{Additional Experiments for Placement of Bank Selection Instruction.}

We also test our Placement of the Bank Selection Instruction with benchmark regression test for architecture MCS51 and Z80, and got similar results to HC08 I stated in Chapter 6.

\begin{figure}
	\centering
		\includegraphics[clip,width=.5\textwidth]{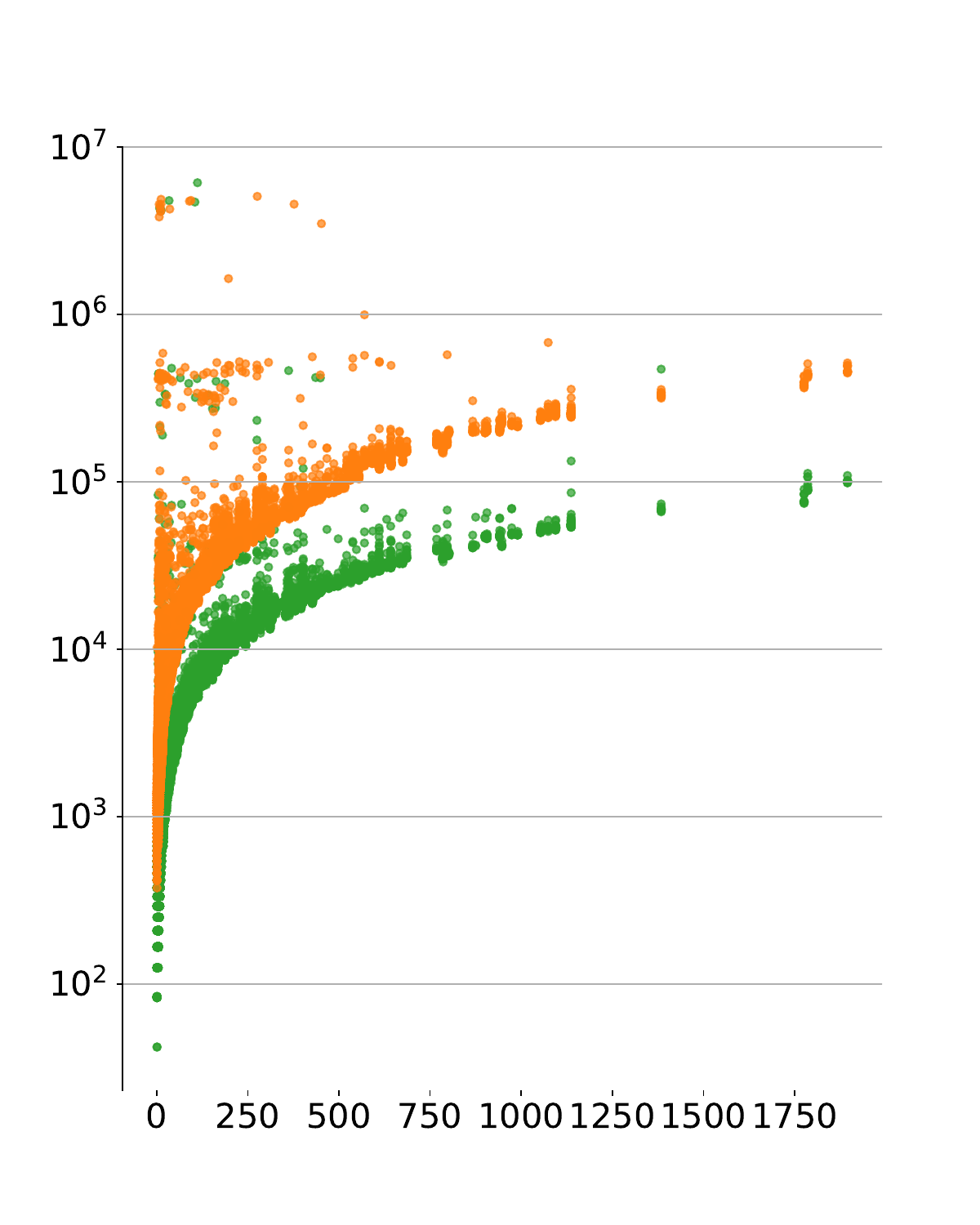}
	
	\caption{Runtime comparison of our approach with the treewidth-based algorithm \cite{bank_selection} over Z80.}
	\label{fig:runtime-f3}
\end{figure}
\newpage

\chapter{Conclusion}
\label{chp_background}
In this work, We presented a general efficient parametric algorithm for binary relationship PCSPs based on SPL-decomposition. We demonstrated that our solution can be applied to various compiler optimization tasks that utilize control flow graphs, including register allocation \cite{RegisterAllocation}, Lifetime-optimal Speculative Partial Redundancy Elimination (LOSPRE) \cite{LOSPRE}, and the placement of bank selection instructions \cite{bank_selection}. The experimental results indicate that our algorithms show significant improvements for all three tasks compared to the previous state-of-the-art algorithms in practice. Furthermore, it is reasonable to assume that for other compiler optimization tasks currently based on tree decomposition, it would be valuable to explore the application of my algorithm and SPL-decomposition.

\newpage

\addcontentsline{toc}{chapter}{Bibliography}
\bibliographystyle{acm}
\bibliography{references}
\newpage

%%%%%%%%%%%%%%%%%%%%%%%%%%%%%%%%%%%%%%%%%%%%%%%%%%%%%%%%%%%%%%%%%%%%%%%%%
%                                                                       %
%     10) APPENDIX (If Any)                                              %
%                                                                       %
% \appendix command marks the beginning of the APPENDIX part of the     %
% Thesis. The usual \chapter command is used for the different chapters %
% of the Appendix.                                                      %
%                                                                       %
%%%%%%%%%%%%%%%%%%%%%%%%%%%%%%%%%%%%%%%%%%%%%%%%%%%%%%%%%%%%%%%%%%%%%%%%%

%%%%%%%%%%%%%%%%%%%%%%%%%%%%%%%%%%%%%%%%%%%%%%%%%%%%%%%%%%%%%%%%%%%%%%%%%
%                                                                       %
%     11) BIOGRAPHY (optional)                                          %
%                                                                       %
% \biography and \endbiography are used to define the optional          %
% Biography of the author of the Thesis.                                %
%                                                                       %
%%%%%%%%%%%%%%%%%%%%%%%%%%%%%%%%%%%%%%%%%%%%%%%%%%%%%%%%%%%%%%%%%%%%%%%%%

\end{document}